\documentclass[aps,prd,showpacs,twocolumn,superscriptaddress,nofootinbib,preprintnumbers]{revtex4-1} 

\makeatletter
\def\p@subsection{}
\makeatother

\usepackage{graphicx,amsmath,amssymb,lipsum,bm}

\usepackage[usenames,dvipsnames]{xcolor}
\definecolor{xlinkcolor}{rgb}{0.7752941176470588, 0.22078431372549023, 0.2262745098039215}

\usepackage[colorlinks=true,citecolor=xlinkcolor,linkcolor=xlinkcolor,urlcolor=xlinkcolor, backref=false,pdfborder={0 0 0}]{hyperref}

\usepackage[normalem]{ulem}
\usepackage[utf8]{inputenc} 
\usepackage{mathtools}
\usepackage{aas_macros}
\usepackage{float}

\usepackage{ulem}
\usepackage{diagbox}

\usepackage[dvipsnames]{xcolor}
\usepackage{mathrsfs}

\newcommand{\be}{\begin{equation}}
\newcommand{\ee}{\end{equation}}
\newcommand{\beqa}{\begin{eqnarray}}
\newcommand{\eeqa}{\end{eqnarray}}

\def\cGpc{\, h^{-3} \, {\rm Gpc}^3}

\renewcommand\k{{\bm k}}
\newcommand\q{\bm{q}}

\newcommand{\bseq}{\begin{subequations}}
\newcommand{\eseq}{\end{subequations}}

\def\gsim{\raise0.3ex\hbox{$\;>$\kern-0.75em\raise-1.1ex\hbox{$\sim\;$}}}
\def\lsim{\raise0.3ex\hbox{$\;<$\kern-0.75em\raise-1.1ex\hbox{$\sim\;$}}}

\def\beqn#1{\begin{equation}\label{#1}}
\def\eeqn{\end{equation}}

\def\beqa#1{\begin{eqnarray}\label{#1}}
\def\eeqa{\end{eqnarray}}

\def\kmax{{k_\text{max}}}
\def\hMpc{h{\text{Mpc}}^{-1}}

\def\Z2{$\mathcal{Z_2}$}

\newcommand {\ignore}[1]{}

\begin{document}

\preprint{MIT-CTP/6019}

\title{Cosmic Shear in Effective Field Theory at Two-Loop Order:\\
Revisiting $S_8$ in Dark Energy Survey Data
}

\author{Shi-Fan Chen}
\email{sc5888@columbia.edu}
\affiliation{Department of Physics, Columbia University, New York, NY 10027, USA}
\affiliation{NASA Hubble Fellowship Program, Einstein Fellow}

\author{Joseph DeRose}
\email{ jderose@bnl.gov}
\affiliation{Physics Department, Brookhaven National Laboratory, Upton, NY 11973, USA}
\affiliation{Lawrence Berkeley National Laboratory, 1 Cyclotron Road, Berkeley, CA 94720, USA}

\author{Mikhail M. Ivanov}
\email{ivanov99@mit.edu}
\affiliation{Center for Theoretical Physics -- a Leinweber Institute, Massachusetts Institute of Technology, 
Cambridge, MA 02139, USA} 
 \affiliation{The NSF AI Institute for Artificial Intelligence and Fundamental Interactions, Cambridge, MA 02139, USA}

\author{Oliver H.\,E. Philcox}
\email{ohep2@cantab.ac.uk}
\affiliation{Leinweber Institute for Theoretical Physics at Stanford, 382 Via Pueblo, Stanford, CA 94305, USA}
\affiliation{Kavli Institute for Particle Astrophysics and Cosmology, 382 Via Pueblo, Stanford, CA 94305, USA}

\begin{abstract} 
\noindent Cosmic shear is a powerful probe of cosmological distances, matter abundance and clustering in the low-redshift Universe.
% Cosmic shear probes 
% small scales where linear cosmological
% perturbation theory breaks down, necessitating careful modeling of nonlinearities. 
Cosmological 
parameter extraction from cosmic shear data is limited by our understanding of baryonic astrophysics, which severely restricts the range of scales used 
in such analyses. We show that the remaining scales are largely perturbative and can be accurately described with
two-loop effective field theory 
(EFT)
predictions. 
We present the first consistent 
analysis of the public 
cosmic shear data from the DES-Y3 catalogs in EFT at the two-loop order, renormalizing small-scale sensitivity in cosmic-shear predictions via a lensing-counterterm expansion and accounting for the intrinsic alignments of galaxies with spin-2 EFT predictions. We constrain the lensing amplitude competitively with standard (empirically-modeled) methods, finding $S_8 = 0.783^{+0.038}_{-0.031}$ ($S_8 = 0.802^{+0.031}_{-0.026}$ with BAO). The perturbativity of cosmic shear suggests novel opportunities for testing new physics with ongoing and upcoming cosmic shear experiments like Roman, Euclid, and LSST. As an example, we derive matter clustering constraints within the dynamical dark energy model from a combination of our DES-EFT cosmic shear likelihood, early-universe CMB priors, DESI BAO, and supernovae data,  finding $S_8 = 0.824\pm 0.029$,
indicating no $S_8$ tension in
the growth of cosmic structure regardless of the underlying cosmological model and expansion history.
\end{abstract}

\maketitle

\textit{Introduction. ---} 
Cosmic shear, the coherent distortion of the apparent shapes of distant galaxies due to gravitational lensing induced by large-scale structure, is one of the leading probes
of cosmological parameters, providing some of the most
competitive constraints on the matter abundance, cosmological distances and mass fluctuations
in the late Universe. These constraints provide key insights into the nature of dark energy,
dark matter, and the values of neutrino masses. 
The significance of cosmic shear 
is emphasized by an array of completed (KiDS~\cite{Wright:2025xka}, 
Dark Energy Survey (DES)~\cite{DES:2026mkc}), 
ongoing (HSC~\cite{Miyatake:2023njf}, Euclid~\cite{Amendola18}) and upcoming surveys 
(Roman~\cite{2019arXiv190205569A}, LSST~\cite{LSST}).

The parameter best constrained by cosmic shear is the lensing amplitude $S_8\equiv\sigma_8\sqrt{\Omega_m/0.3}$, where
$\sigma_8$ is the linear mass fluctuation amplitude at redshift $z=0$ and $\Omega_m$ is the current energy fraction of matter in our Universe.
Historically, measurements of $S_8$ have been in some tension with the CMB-based cosmological standard model $\Lambda$CDM. The most recent 
constraints from the DES-Y6 shear catalogs $S_8=0.798^{+0.014}_{-0.015}$ 
are $\approx 2\sigma$ lower than the 
Planck CMB prediction
$S_8=0.832^{+0.013}_{-0.013}$~\cite{Planck:2018vyg}, while previous results from DES-Y3
$S_8=0.759^{+0.025}_{-0.023}$~\cite{DES:2021bvc}
were lower by about $3\sigma$.
This discrepancy, known as 
the $S_8$ tension, is often interpreted as a potential sign of new physics (see~\cite{Abdalla:2022yfr} for a review).

Unlike the CMB, cosmic shear data cannot be described by linear cosmological perturbation theory. 
Even in the absence of survey systematics, the observed shear is shaped by a variety of non-linear effects:
the collapse 
of matter into halos, baryonic feedback, and the intrinsic alignments (IAs)
of galaxies, which act as a contamination of the weak lensing signal~\cite{Hirata:2004gc}.

The standard approach to modeling non-linearities due to matter clustering 
in cosmic shear
relies on numerical modeling 
by means of N-body simulations. The output of these simulations
over a grid of cosmological parameters is then used to construct 
fitting functions (e.g. \texttt{Halofit}~\cite{Smith:2002dz}, \texttt{HMCode}~\cite{Mead:2020vgs}) or 
emulators, which interpolate over the simulation data (e.g.~\texttt{Aemulus}-$\nu$~\cite{DeRose:2023dmk}).
These approaches typically accurate models for the dark matter distribution at the few to one percent level, with biases from baryonic effects (not captured by N-body simulations) typically mitigated by using scale 
cuts informed by hydrodynamical simulations or marginalized over with phenomenological parameters.
In practice, this restricts many cosmic shear analysis
to comoving wavenumbers $k\lesssim 1~\hMpc$~\cite{DES:2021bvc,DES:2021vln,DES:2022qpf,DES:2026mkc}, though current surveys also run up against limitations from shape noise at similar scales.

The increasing precision of 
future weak lensing data poses two key challenges. 
First, given sub-percent statistical errors, how can we ensure that cosmological constraints are 
robust with respect to theoretical uncertainties due to non-linear clustering and baryonic feedback? 
Second, given the high cost of running numerical simulations,
how can we generalize cosmic shear analyses to physical scenarios beyond the standard model?

In this \textit{Letter} we address these questions by 
developing a consistent treatment of cosmic shear statistics within the effective field theory for large scale structure (EFT)~\cite{BNSZ12,CHS12,Ivanov:2022mrd}
at two-loop order~\cite{Carrasco:2013mua,Foreman:2015lca,Baldauf:2015aha}, which provides a non-linear extension of standard cosmological perturbation theory.
While it has not been the subject of much study in the context of weak lensing (though see Refs.~\cite{DAmico25,Saraivanov:2026sxc}), this approach
has become standard
in analyses of redshift-space galaxy clustering data~\cite{Ivanov:2019pdj,DAmico:2019fhj,Chen:2021wdi,DESI:2024hhd,Chudaykin:2025aux}. 
EFT calculations have also been consistently extended to other tracers,
including projected clustering~\cite{DES:2020yyz,DES:2021zxv,Giri:2023ghr,Chen:2024vvk,Maus:2025rvz,Ivanov:2026dvl}, CMB lensing~\cite{Ivanov:2019hqk,
Chen:2022jzq,Chen:2024vuf}, the Lyman-$\alpha$
forest~\cite{Chen:2021rnb,Ivanov:2023yla,Ivanov:2024jtl} and peculiar velocities \cite{Chen26}.

The key advantage of EFT over simulation-based
techniques is that it allows for a fully analytic treatment 
of non-linearity, and hence can be easily extended to
cosmological models beyond $\Lambda$CDM (as has been demonstrated
in many papers on galaxy clustering, e.g.~\cite{Ivanov:2020ril,Xu:2021rwg,Rogers:2023ezo,Chudaykin25}). 
In addition,
EFT provides 
a systematic way to increase the precision 
of the calculation to match the  target error bars of data, through the inclusion of higher-order corrections (\textit{i.e.}\ additional loops). 

Crucially, two-loop EFT
provides a sub-percent level accuracy
in the description of wavenumbers 
\mbox{$k\lesssim 1~\hMpc$ (at $z\sim 1$)},
which match the scales used in 
cosmic shear analyses. This implies that the scales which are robust with respect to baryonic feedback uncertainties are perturbative,
suggesting that the use of EFT, in principle, should not lead to an significant 
loss of constraining power. In this work we 
explicitly demonstrate that this is indeed the case.

\begin{figure}
    \centering
\includegraphics[width=1.0\linewidth]{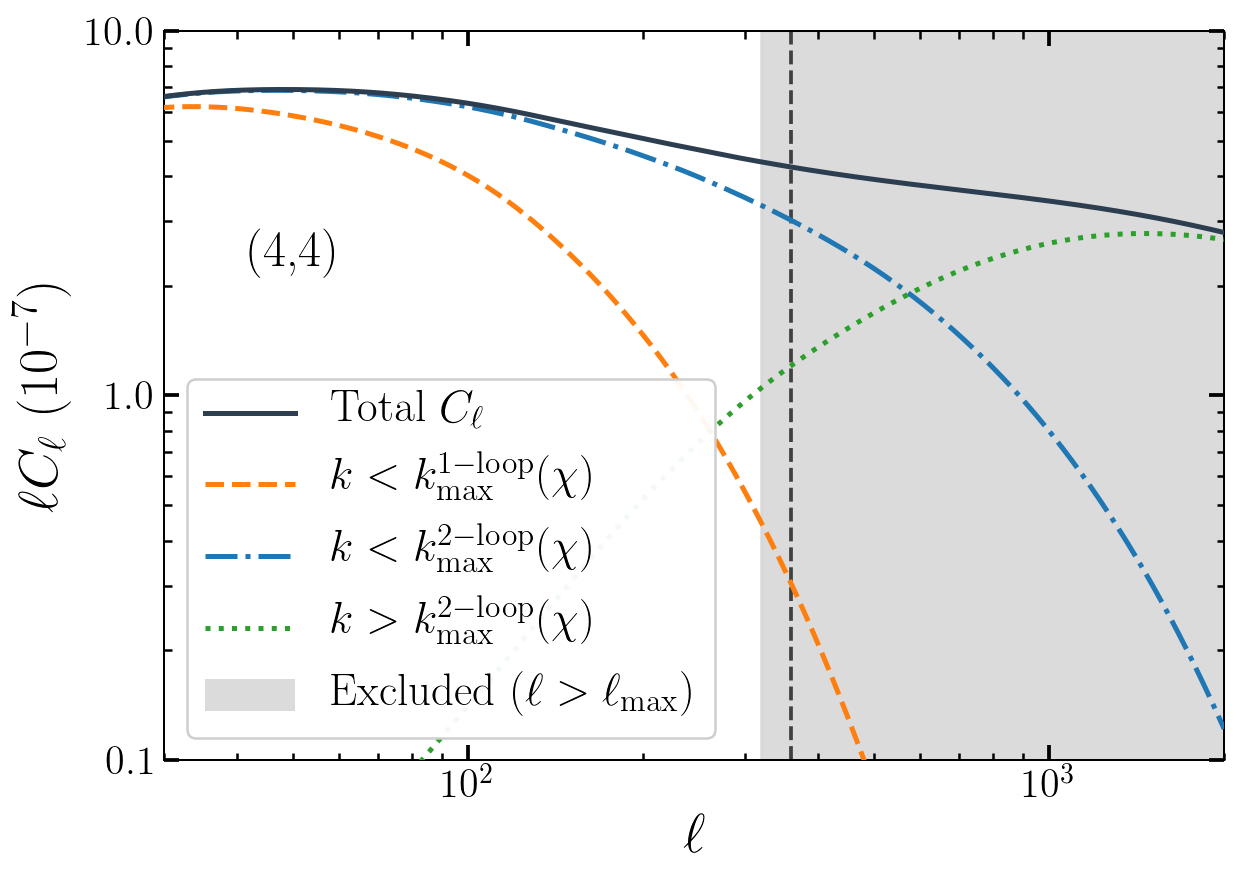}
    \caption{Contribution of different modes to the power spectrum of weak lensing convergence from bin 4 of DES-Y3. The total non-linear signal (black) is split between
    modes captured by one-loop 
     ($k<k_{\rm max}^{\rm 1-loop}(\chi)$) and two-loop EFT ($k<k_{\rm max}^{\rm 2-loop}(\chi)$). 
     Two-loop EFT captures the bulk of the lensing signal for angular scales $\ell< \ell_{\rm max}$, where $\ell_{\rm max}=319$ (for bin 4) used in this study is similar to that commonly used in weak lensing analyses based on empirical modeling, 
     e.g.~\cite{DES:2022qpf} ($\ell_{\rm max}= 360$, vertical dashed line).
     Non-perturbative modes (green dotted) contribute $\lesssim 20\%$ of the total signal at $\ell<\ell_{\rm max}$; in our work, these are marginalized over using lensing counterterms.}
    \label{fig:44_decom}
\end{figure}

In our work, we overcome several crucial challenges in a consistent application
of two-loop EFT to cosmic shear data.
First, we design an accurate and efficient 
emulator for the two-loop theory
computations, leveraging a number of physical scalings. Second, we adapt and upgrade the methodology
of lensing counterterms to marginalize over small-scale uncertainties~\cite{DeRose25}.
Third, we take a model-agnostic approach to EFT nuisance parameters capturing small-scale non-linearities
and baryonic feedback.

\textit{Estimates. ---}  
To motivate the two-loop EFT in the context of cosmic shear, let us discuss the structure of the harmonic-space lensing power spectrum (under the Limber approximation~\cite{Limber:1953}):
\be \label{eq:Cell}
C^{ij}_\ell = \int d\chi\frac{W^i(\chi)W^j(\chi)}{\chi^2}P_{\rm m}\left(k=\frac{\ell+1/2}{\chi},z(\chi)\right)\,,
\ee 
where $\chi$ is the co-moving distance, $\ell\sim \pi/\theta$
is harmonic multipole moment (with corresponding angular correlation scale $\theta$),  
$P_{\rm m}(k,z)$
is the 3D real-space
matter power
spectrum and $W^{i,j}$
are lensing convergence kernels
from sources in bins $i,j$. 
For concreteness, we focus on the 
high-redshift source bin of
DES-Y3 shear data, which peaks at $z_p\approx 0.4$~\cite{DES:2022qpf}.

Perturbation theory (EFT) computations of $P_{\rm m}$
at a given order are reliable up to a certain momentum cut 
$k_{\rm max}$. Previous studies suggest 
that the commonly available 
one-loop computation works up to
$\kmax\approx 0.2~\hMpc$~\cite{Foreman:2015lca,Baldauf:2015aha,Chudaykin:2020hbf}
at $z_p$, which translates
into $\ell_{\rm max}\approx 200$ (or equivalently, $\theta_{\rm min}\simeq 50'$). For larger $\ell$, EFT cannot accurately model the datavector, thus it loses its advantages over emulators. Hence, the bulk of the weak lensing signal remains inaccessible with 
the one-loop EFT.
However, the two-loop EFT computation at $z_p$
is valid for 
$\kmax\approx 0.6~\hMpc$~\cite{ForSen16,Foreman:2015lca,Bal15,Baldauf:2015tla,Baldauf:2015aha},
which translates into $\ell_{\rm max}\approx 700$ ($\theta_{\rm min}\simeq 15'$). This covers almost the entire range of scales normally utilized 
in weak lensing analyses~\cite{DES:2021vln,DES:2022qpf}. 

While the above estimate is quite crude, it nevertheless suggests that a significant fraction of weak lensing data can be analyzed analytically using EFT methods.
In the Supplemental Material we carry out a more careful 
estimation including the width of 
the lensing kernel, which confirms that 
modes captured by EFT at two-loops
indeed account for the bulk of the lensing signal used in cosmological 
analyses. Fig.~\ref{fig:44_decom} shows an example of this in the case of the high-redshift DES-Y3 source sample. While weak-lensing observables are indeed sensitive to small-scale physics (as emphasized in several recent works, e.g.~\cite{Amon22}), this effect is due to the finite extent of the lensing kernel to small distances ($z<z_p$) and can be renormalized by lensing counterterms, rather than contribute with arbitrary spectral shapes \cite{DeRose25}, as we discuss below. 

\textit{Theory description. ---}  
At two-loop order, the EFT description of the matter power spectrum (which describes cosmic shear) can be written~\cite{CHS12,Carrasco:2013mua,Foreman:2015lca,Baldauf:2015aha,Bakx26}:
\begin{widetext}
\begin{align}
    P_{\rm m}(k,z) &= P_{\rm m}^{\rm lin}(k,z) + P_{\rm m}^{\rm 1-loop,~SPT} (k,z) + P_{\rm m}^{\rm 2-loop~SPT}(k,z) - 2 c^2_s(z) \left( \frac{k}{k_{\rm nl}(z)} \right)^2 P_{\rm m}^{\rm lin}(k,z) \nonumber \\
    &\quad  - 2 c^2_{s,1}(z)  \left( \frac{k}{k_{\rm nl}(z)} \right)^2 P_{\rm m}^{\rm 1-loop}(k,z)  - 2 c_4(z)  \left( \frac{k}{k_{\rm nl}(z)} \right)^4   P_{\rm m}^{\rm lin}(k,z)  - 2  c_{\rm quad}(z)  \left( \frac{k}{k_{\rm nl}(z)} \right)^2  P_{\rm m}^{\rm quad}(k,z) 
    % \nonumber\\
    % &\quad  
    \,,
    \label{eqn:p2loop}
\end{align}
\end{widetext}
where 
the first line represents 
the two-loop matter contributions
in standard Eulerian perturbation
theory (SPT)~\cite{Ber02,Blas:2013bpa}, with the remaining terms describing higher-derivative EFT corrections that renormalize the contributions of non-perturbative small-scale modes. The parameters
$c^2_s,c^2_{s,1},c_4,c_{\rm quad},c_\epsilon$ above are 
redshift-dependent EFT
parameters (counterterms) and
$P_{\rm m}^{\rm quad}$
is a one-loop power spectrum built
from the quadratic derivative 
operator $\sim (\nabla^2\delta^2_1)$
($\delta_1$ is the linear overdensity). 
In principle, there are four distinct quadratic operators~\cite{Baldauf14,Konstandin19,Steele:2020tak,Baldauf:2021zlt,Anastasiou25}; but since their contributions to the power spectrum are highly degenerate, we keep only one such term in this \textit{Letter} (see Supplemental Material for
more detail).\footnote{These contributions, however, can be distinguished at the level of the one-loop bispectrum~\cite{Steele:2020tak}.}
We also find that other 
cubic and stochastic terms are
degenerate with other EFT counterterms,
which motivates using
a minimal basis of only
four parameters
$\{c^2_s,c^2_{s,1},c_4,c_{\rm quad}\}$ 
in our two-loop theory prediction.
In addition we implement IR resummation to account for the
non-linear distortions of baryon acoustic oscillations (BAO) following Refs.~\cite{SenZal15,Baldauf15,Blas16,Vlah16,Ivanov18}.

The normalization scale in Eq.~\eqref{eqn:p2loop}
is 
the non-linear scale $k_{\rm nl}(z)$ 
defined as
\begin{equation}
    \frac{k_{\rm nl}^3(z) P_{\rm m}^{\rm lin}(k_{\rm nl}(z),z)}{2\pi^2} = 1\,,
\end{equation}
which is a proxy for the EFT cutoff.
% $\Lambda_{\rm EFT}$. 
Notably, normalizing EFT counterterms via the time-dependent scale $k_{\rm nl}$ absorbs much of their redshift evolution. In a power-law universe 
with $P_{\rm m}^{\rm lin}\propto k^n/k_{\rm nl}^{3+n}$, this choice of normalization would exactly capture the redshift-dependence~\cite{PajZal13}; in our Universe, however, the scaling is approximate and does not, for example, exactly capture the cutoff-dependence of the loop integrals in the first line of Eq.~\eqref{eqn:p2loop}. 

The $k_{\rm nl}(z)$ rescaling enables us to model the time evolution of the counterterms via smoothly varying order-one functions of redshift even while the dimensionful counterterms vary by more than an order of magnitude over our redshift range. Specifically,
we follow~\cite{Chen24BAO,Chen24,DeRose25} (see also Ref.~\cite{Saraivanov:2026sxc}) and model the time dependence of EFT parameters using a spline basis:
\begin{equation}
\label{eqn:spline_time_dep}
\ c_{O^{(n)}}(z) = \sum_{i=1}^N c_{O^{(n)}}^i W_s\left(\frac{z - z_{\rm min}}{\Delta} - \left(i - \frac{s-1}{2} \right)\right)
\end{equation}
where $N$ is the number of spline nodes, $n$ is the leading order of the operator and $W_s$ is the spline kernel of order $s$, which we here set to two \cite{Hockney88,Jeong10}.\footnote{Here $W_0(x) = \Theta( \frac12 - |x|)$ and $W_s = W_0^{\ast (s+1)}$ are convolutions.} In addition, we rescale the EFT parameters by appropriate powers of $\sigma_8(z)/\sigma_{8,{\rm fid}}$
in order to suppress prior-volume effects~\cite{Maus25,Chen:2024vvk,Tsedrik25,Chudaykin25}.

We emphasize that EFT is a
data-driven approach, \textit{i.e.}\ in principle,
one does not require any calibration from simulations. 
In particular, one can 
model the time-dependence
of the EFT parameters purely based on theoretical arguments and the data itself. 
This is the spirit of our spline model, which is flexible enough to capture a variety of time-dependencies.
In the Supplemental Material we show that by using a quadratic spline with $N=4$,  totaling 16 free EFT parameters, we can fit matter power spectra 
from the $\textsc{Aemulus}-\nu$
simulations to $0.5\%$ on quasi-linear scales, with comparable results to fitting each redshift independently; this reflects the smooth evolution of non-linear structure \cite{Foreman:2015lca}. We note that our $S_8$ constraint is rather insensitive to the choice of $N$.
In the Supplemental Material, we additionally show that counterterms due to baryonic effects~\cite{Lewandowski15,Braganca:2020nhv} are degenerate with, or indeed captured by, a subset of the two-loop EFT counterterms, and do not need to be treated independently in our analysis. We also discuss the treatment of intrinsic alignments with EFT, which we model within the Lagrangian EFT of intrinsic alignments \cite{Vlah20,Taruya21,Chen23} following Refs.~\cite{Chen24,DeRose25}.

\textit{Lensing counterterms.---}
A crucial feature of our EFT methodology is a consistent treatment of small-scale physics, and its impact on large-scale modes. Even though the 3D power spectrum is renormalized, its projection integral over the lensing kernel in Eq.~\eqref{eq:Cell} still receives
significant contributions
from small-scale modes in the limit $\chi\to 0$. This extra ultraviolet (UV) sensitivity is mitigated by \textit{lensing counterterms},
which are computed by Taylor expanding the lensing integrand in the UV, resulting in a series of free EFT coefficients capturing the matter power spectrum on UV scales \cite{DeRose25}. 
We estimate
the theoretical error due to such modes using N-body predictions from $\textsc{Aemulus}-\nu$.
One conceptual feature of EFT 
is the time-dependence of the
cutoff in Eq.~\eqref{eq:Cell}, since $k_{\rm max}\sim k_{\rm nl}(z)$ moves to smaller scales at earlier times,  which is described in detail in Supplemental Material. Lensing counterterms add up six free parameters $f_{No}^{\rm UV}$ starting with the leading one at $N, o = 2, 0$.

\textit{Two-loop  theory emulation. ---}   The two-loop
theory predictions
depend on the two-loop integral 
$P_{\rm m}^{\rm 2-loop,~SPT}$ for 
which efficient numerical 
methods such as
FFTLog~\citep{Simonovic:2017mhp,Chudaykin:2020aoj} are not currently available. 
To overcome this bottleneck, 
we produce an emulator trained on numerical calculations of the loop integrals, themselves obtained using Monte Carlo methods \cite{Blas:2013bpa,Blas:2013aba,Konstandin19}.
To capture the 
cosmology-dependence of the power spectrum shape, we evaluate $P_{\rm m}^{\rm 2-loop}$  
for a wide range of $\omega_{\rm cdm}$ values. The dependence on $h$, dark energy
equation of state parameters $w_0,w_a$
and primordial amplitude $A_s$
is taken into account exactly via rescalings of the momentum $k$ and the linear power spectrum normalization
including late-universe modifications to the growth factor.
We do not currently vary the baryon density $\omega_b$ or power spectrum tilt $n_s$ since weak lensing data are not sensitive to these parameters. This justifies using strong CMB-based priors on $\omega_b$ and $n_s$, or, more practically, fixing them to the Planck 2018 best-fit values, which we adopt throughout~\citep{Planck:2018vyg}. It would be straightforward to include these (and any other) parameters in the emulator if needed.

\textit{Application to data. ---} 
As a first application of our theoretical EFT model, we analyze the cosmic shear power spectra presented in Ref.~\cite{DeRose25}. These data are based on the Dark Energy Survey Year 3 (DES-Y3) shape catalog \cite{DESY3Catalog} presented in Ref.~\cite{DeRose25} and obtained using the \texttt{NaMaster} power spectrum estimator \cite{namaster} with Gaussian covariances derived from the improved narrow kernel approximation \cite{Nicola21}.

To validate our analysis choices, in particular regarding the modeling of non-linearities, we test our pipeline on mock data vectors produced using fully non-linear power spectra predicted from the $\textsc{Aemulus}-\nu$ emulator. This was shown to reproduce cosmic-shear measurements from the \texttt{Buzzard} mocks (which were used to validate DES analyses \cite{Buzzard}) within their statistical precision \cite{DeRose25}. These mock data vectors are noiseless but fully non-linear realizations of DES observables; we show that applying our model to them yields unbiased cosmological constraints given our various analysis choices (see the Supplemental Material). We explore two particular fitting settings, a Conservative Cut with $k_{\rm max}$ determined by fitting to $k_{\rm nl}$, and an Empirical Cut determined by fitting to simulations, finding unbiased constraints and theoretical predictions under theoretical control in both cases, motivating us to set the latter as our default.

\begin{figure}
    \centering
    \includegraphics[width=1.0\linewidth]{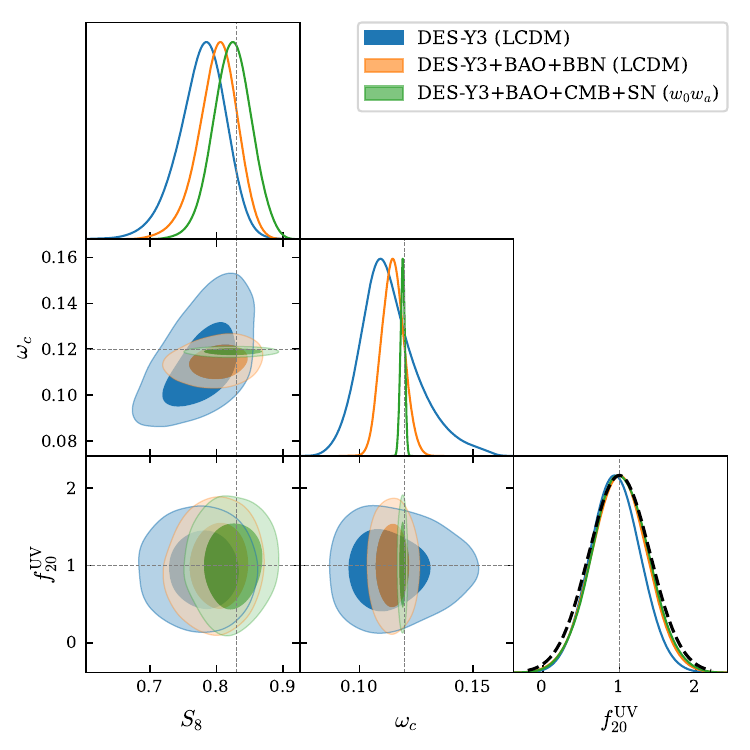}
    \caption{Cosmological constraints from applying our two-loop EFT model to DES-Y3 data in $\Lambda$CDM (w/ and w/o DESI-DR2 BAO and BBN) and $w_0 w_a$CDM (combining with DESI-DR2 BAO, late-universe marginalized Planck CMB, and SNIa from DES-Y5). Dashed solid and vertical lines mark the best-fit $\Lambda$CDM cosmology from the Planck CMB. The dashed contours show our priors on the leading lensing counterterm, which the data only loosely constrain.}
    \label{fig:constraints}
\end{figure}

Turning to the results, we find that $S_8 = 0.781^{+0.043}_{-0.035}$ (Conservative Cut) and $S_8 = 0.778^{+0.038}_{-0.031}$ (Empirical Cut) within $\Lambda$CDM. In line with our estimates on the sizes of loop terms, these constraints are quite competitive with the official DES-Y3 harmonic space constraints~\cite{DES:2022qpf} ($S_8 = 0.793^{+0.038}_{-0.025}$) even though the latter do not allow for baryonic or dynamical non-linearites beyond CDM. 
Our $S_8$ agrees well with 
Planck $\Lambda$CDM value $S_8=0.832\pm 0.013$, 
which can be contrasted with the 
DES-Y3 configuration space result
$S_8=0.759^{+0.025}_{-0.023}$~\cite{DES:2021bvc}.
Note that weakening the \textit{physical} scale cuts does not dramatically worsen the constraints since shorter scales primarily constrain EFT parameters (similar to how small-scales primarily constrain feedback parameters in conventional analyses \cite{DES:2022qpf,DeRose25}).

To investigate the applicability of our method beyond the standard $\Lambda$CDM paradigm we analyze the DES data allowing for dark energy beyond the cosmological constant within the $w_0 w_a$ parametrization \cite{CPL}. 
A consistent non-linear analysis of this model was not previously possible since the commonly used emulators do not provide a sufficient coverage of the parameter space of this model~\cite{Mead:2020vgs}.

Cosmic shear is sensitive to modifications in the expansion history through the lensing kernel, and recent analyses combining BAO from the DESI survey, CMB and supernovae data show a mild preference for $w_0, w_a$ beyond $\Lambda$CDM \cite{DESI-DR2-BAO}. 
To test its impact on the $S_8$ inference, we obtain constraints within a $w_0 w_a$ cosmology by combining the DES-Y3 data with DESI-DR2 BAO, DES-Y5 supernovae \cite{DESY5} and CMB data from Planck with late-universe effects (e.g. lensing and integrated Sachs-Wolfe effect) marginalized away \cite{Lemos23}.\footnote{Specifically, we train a normalizing flow using \texttt{FlowJax} on the chains run with these data in Ref.~\cite{DESI-DR2-BAO} and use it to emulate the corresponding likelihood.} The latter choice means that we only use CMB information on the geometry and power spectrum shape, such that the constraint on the clustering amplitude is derived from cosmic shear alone. 
With these external priors, the DES-Y3 data constrain $S_8 = 0.824\pm 0.029$. The tighter error bars reflect that these additional datasets tightly constrain low-redshift distances and the shape of the power spectrum, breaking residual degeneracies of other cosmological distances with $S_8$.
Indeed, within $\Lambda$CDM, combining the DES-Y3 data with a Big-Bang nucleosynthesis (BBN) prior on $\omega_b$ and DESI BAO data yields $S_8 = 0.802^{+0.031}_{-0.026}$, recovering similarly tight constraints with a slightly shifted $S_8$ due to small shifts in the matter density.

Importantly, the above $S_8$ inferences
are consistent with CMB ones both within 
$\Lambda$CDM ($S_8=0.832\pm 0.013$) and 
$w_0w_a$CDM
models ($S_8=0.827\pm 0.010$, combined with BAO and SNe), implying no $S_8$ tension in the DES-Y3 weak lensing data.
This conclusion
% the coabsence of the $S_8$ tension
is robust w.r.t. assumptions
about the underlying cosmological
model.

A summary of our cosmological constraints is shown in Figure~\ref{fig:constraints}. The best-fit model within in the $w_0 w_a$CDM cosmology is shown in the Supplementary Materials, where we also show the relative importance of the EFT formalism and lensing counterterms by plotting model predictions without either ingredient. Our results agree well with CMB and also with the EFT-based 
measurements of $S_8$ from spectroscopic clustering, both in $\Lambda$CDM and $w_0w_a$CDM cosmologies~\cite{DESI:2024hhd,Maus:2025rvz,Chudaykin:2025aux,Ivanov:2026dvl,Chudaykin:2026nls}, and indicate that the DES 
shear data is in agreement with 
other probes of cosmic structure. The inclusion of DES-Y3 data in combined analyses does not lead to significantly sharper constraints on $w_0, w_a$, since the distances in the lensing kernel are already fixed to sub-percent level by the other probes (CMB, BAO, and supernovae) -- we intend to return to this topic with future lensing samples as well as the full CMB dataset.

\textit{Conclusions. ---}  
In this \textit{Letter} we have developed a consistent formalism to analyze cosmic shear power spectra using EFT at two-loop order. An important conclusion is that much of the data used in standard shear analyses can be robustly described by perturbative models (after marginalizing over baryonic physics uncertainties); this opens up a range of exciting research directions. 

First, one may extend the cosmic
shear analyses to other beyond-$\Lambda$CDM scenarios, including ultralight axions, early dark energy, and beyond. This will require efficient tools for two-loop computations (particularly as the size of the parameter space increases), such as the \textsc{cobra} and massive propagator basis functions \cite{Bakx25,Bakx26,Anastasiou25}.

Second, our analysis can be extended to galaxy clustering and galaxy-galaxy lensing correlations within the
$3\times 2$ point framework.
This will require developing 
two-loop computations 
for galaxies, see Ref.~\cite{Bakx26b}
for recent progress in this direction. This will be particularly important in order to calibrate the effects of galaxy intrinsic alignments \cite{Chen:2024vvk,DeRose26}.

Third, one could consider extending our approach to configuration-space observables, such as the shear correlation function. This will development of analogous lensing counterterms to account 
for mode-mixing due to both projection effects and the Fourier transform over small-scale modes. This will be important to ensure that theoretical predictions are systematically controlled and that cosmological parameter inference is robust.

Finally, it will be interesting to investigate the cosmology-dependence of two-loop EFT parameters for dark matter, including in the presence of baryons. This will be important 
to potentially inform EFT-based analysis through simulation-based priors along the lines of~\cite{Ivanov:2024hgq,Ivanov:2024xgb,Chudaykin:2026nls}.

\vspace{2pt}
\textit{Note added.} While this paper was being prepared, Ref.~\cite{Saraivanov:2026sxc} appeared with forecasts for Roman based on two-loop EFT calculations in configuration space~\cite{Saraivanov:2026sxc}. While we differ in the treatment of UV sensitivities in the lensing integral (lensing counterterms) and intrinsic alignments (Lagrangian EFT), we arrive upon similar conclusions on the applicability of higher-order perturbation theory to cosmic shear.
% Despite  some differences in our methodologies
% the projected constraint on $S_8$
% for LSST-Y10 from this work match
% well our result. 

\textit{Acknowledgements.} 
We thank Vivian Miranda and Elisabeth Krause for 
useful conversations. Support for this work was provided by NASA through
the NASA Hubble Fellowship grant HST-HF2-51572.001
awarded by the Space Telescope Science Institute, which
is operated by the Association of Universities for Research in Astronomy, Inc., for NASA, under contract
NAS5-26555. 

% \appendix 

% \section{Theory model details}

\bibliographystyle{apsrev4-1} % Tell bibtex which bibliography style to use
\bibliography{main.bib}

% \bibliography{low_redshift} % Tell bibtex which .bib file to use (this one is some example file in TexLive's file tree)

\newpage 
\pagebreak

%%%%%%%%%% Merge with supplemental materials %%%%%%%%%%
%\pagebreak
\widetext
\begin{center}
\textbf{\large Supplemental Material}
\end{center}
%%%%%%%%%% Merge with supplemental materials %%%%%%%%%%
%%%%%%%%%% Prefix a "S" to all equations, figures, tables and reset the counter %%%%%%%%%%
\setcounter{equation}{0}
\setcounter{figure}{0}
\setcounter{table}{0}
\setcounter{page}{1}
\makeatletter
\renewcommand{\theequation}{S\arabic{equation}}
\renewcommand{\thefigure}{S\arabic{figure}}
\renewcommand{\bibnumfmt}[1]{[S#1]}
\renewcommand{\citenumfont}[1]{S#1}
%%%%%%%%%% Prefix a "S" to all equations, figures, tables and reset the counter %%%%%%%%%%

\section{Two-loop computations and their reach}

\textbf{Two-loop EFT.} 
Here we provide some details on the two-loop power spectrum expression in Eq.~\ref{eqn:p2loop}.
Schematically, one can write down contributions to the non-linear matter field as~\cite{Carrasco:2013mua,Foreman:2015lca,Anastasiou25}
\be 
\delta_{\rm NL}=\delta_1+\delta_{\rm SPT}^{\rm 1-loop}+\delta_{\rm SPT}^{\rm 2-loop}+\delta_{1}^{\rm ctr.}+\delta_{2}^{\rm ctr.}+\delta_{3}^{\rm ctr.}+\delta_\epsilon\,,
\ee 
where $\delta_1$ is the linear field, $\delta_{\rm SPT}^{n-{\rm loop}}$
are $n$-loop 
field corrections in standard perturbation theory~\cite{Ber02} and $\delta_\epsilon$ is the stochastic correction~\cite{Carrasco:2013mua,Baldauf14}, giving the power spectrum $\propto c_\epsilon k^4$. The linearly independent first and second order counterterms are given by\footnote{Note that the $F_2^{\alpha\beta}$ contribution of~\cite{Baldauf14} can be recast into
the four
$E_n$ even for an arbitrary time-dependence of the dark matter sound speed. This comes at the expense of 
unobservable shifts of the corresponding EFT parameters.}
\be 
\begin{split}
& \delta_{1}^{\rm ctr.}=  -c_s
\frac{k^2}{k_{\rm nl}^2}
\delta_1-c_4\frac{k^4}{k_{\rm nl}^4}\delta_1\,,\quad 
\delta_{2}^{\rm ctr.}=\frac{1}{2k_{\rm nl}^2}\sum_{n=0}^3 e_n \int_{\q_1}
\int_{\q_2}(2\pi)^3\delta_D^{(3)}(\k-\q_{12})
E_n(\k_1,\k_2)\delta_1(\q_1)\delta_1(\q_2)\,,\\
& E_0(\q_1,\q_2)=k^2F_2
\,,\quad  
E_1=k^2\,,\quad E_2=k^2\left(\frac{(\q_1\cdot\q_2)^2}{q_1^2q_2^2}-1\right)  \,,\quad 
E_3=\frac{(\q_1\cdot\q_2)}{2}\left(
\frac{\k\cdot\q_2}{q_2^2}+
\frac{\k\cdot\q_1}{q_1^2}
\right)\,,
\end{split}
\ee 
where $\int_{\q}\equiv \int\frac{d^3q}{(2\pi)^3}$ and $F_2$ is the standard perturbation
theory (SPT) density kernel~\cite{Ber02}.
The operators multiplying $e_0$
and $c^2_{s}$ both stem from the sound speed term in the effective stress-tensor and hence require
only one free parameter (the sound speed) if the time-dependence 
is specified. This is a standard approach in the literature, e.g.~\cite{Baldauf14,Anastasiou25},
which assumes the EFT parameters to have a simple power-law dependence on the growth factor $D_+(z)$. 
However, the time-dependence
of EFT parameters is \textit{a priori} unknown.
The assumptions about the simple power-law dependence done in the literature imply 
very strong priors in weak lensing analysis. In order to be more general
and data driven, we will be 
agnostic about the time-dependence 
of EFT parameters, in which case
$e_0$ and $c_s$ should be treated
as independent redshift-dependent functions.

\begin{figure}[htb!]
    \centering
    \includegraphics[width=1.0\linewidth]{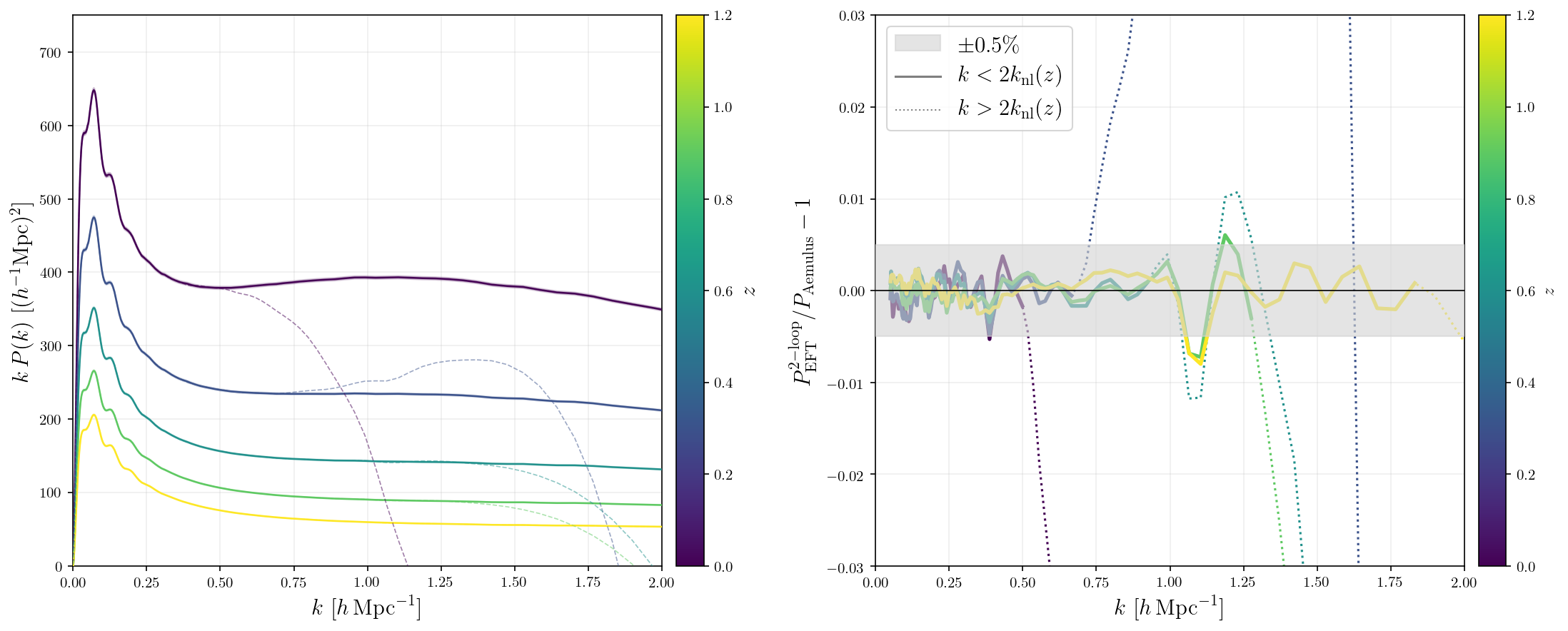}
        \includegraphics[width=1.0\linewidth]{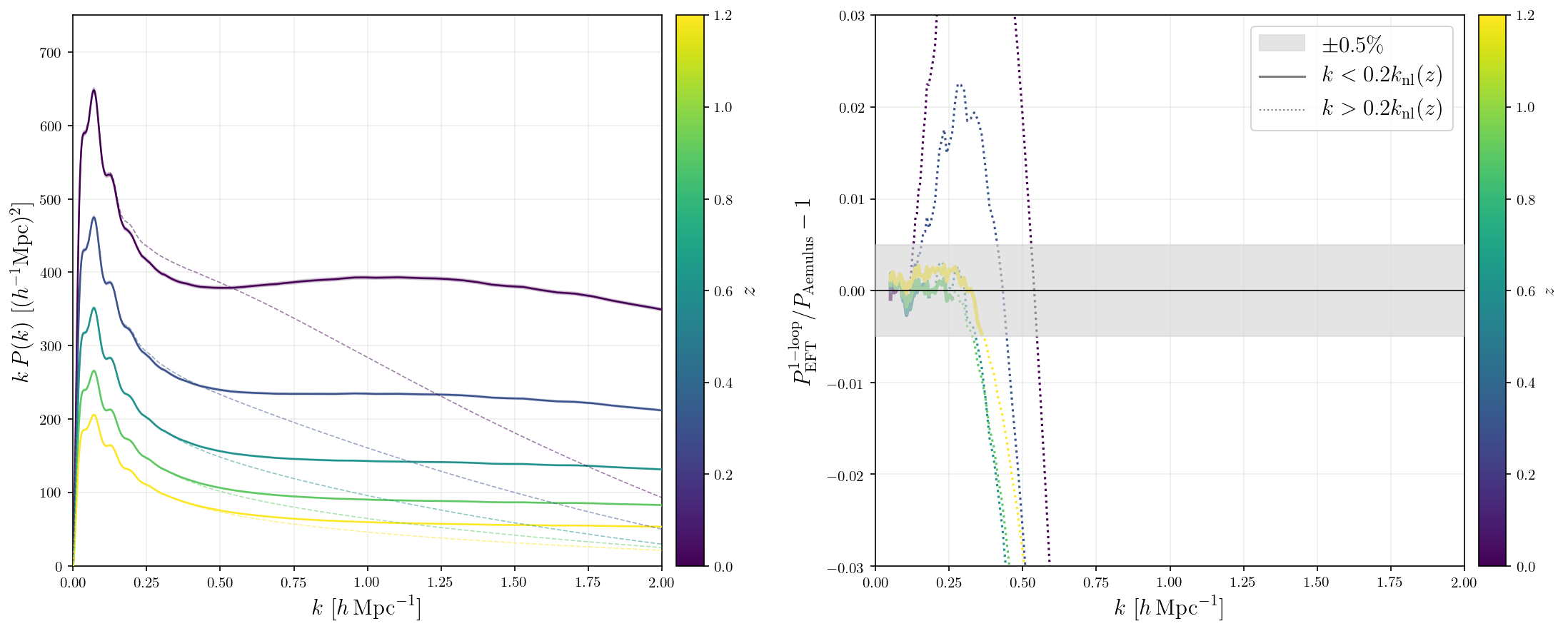}
    \caption{\textit{Left upper panel:} The non-linear matter power spectrum
$P(k)$ at redshifts $z = 0$--$1.2$
(color bar), computed from the $\textsc{Aemulus}-\nu$ emulator. The dashed curves show
the corresponding EFT best fits up to $k_{\rm max}^{\rm 2-loop}=2k_{\rm nl}$, where the 2-loop EFT can reproduce the mock data to $0.5\%$. 
\textit{Right upper panel:} Fractional residual of the two-loop EFT matter power
spectrum relative to the $\textsc{Aemulus}-\nu$ emulator prediction,
$(P^{\rm 2\text{-}loop}/P_{\rm Aemulus} - 1)$ at the
same redshifts. The gray band marks the $\pm 0.5\%$ accuracy threshold. Solid
curves show the residual for modes with $k < 2k_{\rm nl}(z)$ (within the
nominal two-loop validity range) and dotted curves for $k > 2k_{\rm nl}(z)$
(beyond it).
\textit{Lower panels:} As above, but for the one-loop  EFT calculations, with $k_{\rm max}^{\rm 1-loop}=0.2k_{\rm nl}$. This cut is chosen to that the EFT calculation can reproduce the data to the same $0.5\%$ accuracy.}
    \label{fig:aem_fits}
\end{figure}

The cubic counterterms $\delta_{3}^{\rm ctr.}$ have a lengthily expression which we omit here, see refs~\cite{Carrasco:2013mua,Foreman:2015lca,Anastasiou25} for details. These contributions are highly degenerate in practice.
In what follows we will only keep one 
particular cubic term that is needed to cancel the IR enhancements produced by the contraction of $E_0$ and $F_2$ in a $P_{22}$-like diagram. This leads to a counterterm 
$\propto k^2P^{\rm 1-loop}$. 
Computing the power spectrum
of $\delta_{\rm NL}$
and keeping only terms relevant 
at the two-loop order in power counting~\cite{Carrasco:2013mua,PajZal13,Ivanov:2022mrd} yields 
\be 
P^{\rm 2-loop,~EFT}_{\rm m}=P_{\rm m}^{\rm lin}
+P_{\rm m}^{\rm 1-loop,~SPT}+P_{\rm m}^{\rm 2-loop,~SPT}+P_{\rm ctr.}^{\rm 1-loop}+P_{\rm ctr.}^{\rm 2-loop}\,,
\ee 
where $P_{\rm ctr.}^{\rm 1-loop}=-2c_s^2k^2P_{\rm m}^{\rm lin}$ is the one-loop sound speed counterterm. The leading two-loop counterterms 
are given by
\be \label{eq:ctr_2loop}
\begin{split}
 P^{\rm 2-loop}_{\rm ctr.}=&~2\sum_{n=1}^3\frac{e_n}{k_{\rm nl}^2}  \int_{\q} E_n(\k-\q,\q)F_2(-\k+\q,-\q)P_{\rm m}^{\rm lin}(|\k-\q|)
P_{\rm m}^{\rm lin}(q)\\
&~ +2\frac{k^2}{k_{\rm nl}^2} e_0 P^{\rm 1-loop,~SPT}(k)-2c_4\frac{k^4}{k_{\rm nl}^4} P_{\rm m}^{\rm lin}(k)+ \frac{2\pi^2 c_\epsilon(z)}{k_{\rm nl}^3} \left( \frac{k}{k_{\rm nl}} \right)^4\,,
\end{split}
\ee 
where all quantities above are evaluated at redshift $z$.
The two-loop power spectrum is characterized by ${c^2_s,c^2_{s,1},e_1,e_2,e_3,c_4,c_\epsilon}$
at this stage.
We will see shortly that in a realistic
analysis this set of parameters can be
further reduced down to $\{{c^2_s,c^2_{s,1}\equiv e_0,e_1,c_4}\}$, which leads to Eq.~\eqref{eqn:p2loop} with $c_{s,1}^2\equiv e_0$ and $c_{\rm quad}=e_1$:
\be
P_{\rm quad}=2c_{\rm quad}\frac{k^2}{k_{\rm nl}^2}  \int_{\q} F_2(\k-\q,\q)P_{\rm m}^{\rm lin}(|\k-\q|)
P_{\rm m}^{\rm lin}(q)\,.
\ee
Note that one may 
subtract the double-hard limit of $P_{\rm m}^{\rm 2-loop,~SPT}$ proportional to $k^2P_{\rm m}^{\rm lin}$~\cite{Blas:2013bpa,Baldauf:2015aha}, which makes a large cutoff-dependent contribution to the power spectrum degenerate with the one-loop sound speed counterterm. Such subtraction is done, e.g. in fig.~\ref{fig:aem_fits}.
However, this does not make a difference in practice because this
scheme-dependent contribution 
dies off quickly at high redshift.

We implement infrared (IR) resummation via \cite{Blas16,Vlah16}
\be 
P^{\rm IR-res}=P_{\rm nw}+P_{\rm w}e^{-S}\left(1+S+\frac{S^2}{2}\right)+P^{\rm 1-loop}[P_{\rm nw}+P_{\rm w}e^{-S}(1+S)]+P^{\rm 2-loop}[P_{\rm nw}+P_{\rm w}e^{-S}]\,,
\ee 
where $P_{\rm w}$ and $P_{\rm nw}$
are the wiggly and smooth parts of the linear matter power spectrum, and the damping exponent is given by
\be 
S\equiv \frac{k^2}{6\pi^2}\int_0^{\Lambda_{\rm IR}} dq~P^{\rm lin}(q,z)\left[1-j_0(qr_s)+2j_2(qr_s)\right]\,,
\ee 
where $r_s$ is the comoving sound horizon and $j_\ell(x)$
are spherical Bessel functions. 
We tested different choices of the IR cutoff 
$\Lambda_{\rm IR}=0.2~\hMpc$ 
(following~\cite{Blas16,Chen23})
and $\Lambda_{\rm IR}=\infty$ (as in convolutional Lagrangian perturbation theory)
and find consistent results~\cite{Blas16}. The cutoff scale $\Lambda_{\rm IR}=0.2~\hMpc$ is also used to compute one-loop predictions for the intrinsic alignment signals within Lagrangian perturbation theory (as described below) using a similar resummation scheme \cite{Chen21,Chen23}. 

\textbf{Practical Implementation.}
To compute the SPT contributions to the matter power spectrum we use a Monte Carlo integration scheme, implemented using the \textsc{vegas} importance-sampling algorithm. Given an input linear power spectrum (optionally IR-resummed), we compute the one- and two-loop SPT spectra (which involve two- and five-dimensional integrals) across an array of 128 logarithmically spaced points with $k\in[10^{-3},10]\hMpc$. We utilize IR-safe two-loop integrands \citep{Carrasco14} with higher-order kernels computed recursively, with maximal caching. For each $k$-point, we demand that the combined tree-plus-loop integral converges to $<0.1\%$ precision; this ensures that we do not waste time on configurations that 
contribute negligibly to the final result. The computations are performed in \textsc{julia} and are embarrassingly parallelized, requiring a few node-hours per input power spectrum. 

To perform cosmological analyses, we require the power spectrum as a function of cosmological parameters. Here, we can compute $P_{\rm m}^{\rm lin}(k)$ purely as a function of $\omega_{\rm cdm}$ noting that (a) $\omega_b$ and $n_s$ cannot be constrained by weak-lensing datasets, (b) the dependence on $h$ can be fully accounted for via momentum rescalings, and (c) the overall amplitude enters only as a scaling (with $P^{(n-\rm loop)}\propto \sigma_8^{2n}$, up to IR-resummation effects, which we treat separately by computing wiggly and non-wiggly version of the two-loop integrals
following~\cite{Ivanov18}). We compute the power spectrum across a grid of 19 points in $\omega_{\rm cdm}$, with $\Delta \omega_{\rm cdm}=0.05$ spacing for $\omega_{\rm cdm}\in[0.09,0.15]$ and $\Delta\omega_{\rm cdm}=0.1$ spacing for $\omega_{\rm cdm}\in[0.06,0.08]$ and $[0.16,0.18]$. At runtime, the various contributions can be trivially emulated or interpolated.

\textbf{Scale-Dependence and Reach.}
To determine the relevance of various terms, we fit the $\textsc{Aemulus}-\nu$ emulator data for a fiducial dark matter survey with $V=5$ [$h^{-1}$Gpc]$^3$ at redshifts $z=[0,0.3,0.6,0.9,1.2]$ and a Gaussian error covariance as in~\cite{Chudaykin:2020hbf}.
First, empirically, we find that the two-loop calculation with free $\{c^2_s,c^2_{s,1},e_1,e_2,e_3,c_4,c_\epsilon\}$
at each redshift reproduces
the matter power spectrum within $0.5\%$
up to 
$k_{\rm max}^{\rm 2-loop}(z)=2k_{\rm nl}(z)$. This is significantly larger than the reach of the one-loop calculation which we find at the same accuracy, $k_{\rm max}^{\rm 1-loop}(z)=0.2k_{\rm nl}(z)$, consistent with the literature~\cite{Foreman:2015lca}. 
The fact that this scale is larger than
$k_{\rm nl}(z)$ should not be particularly concerning:
$k_{\rm nl}$ is only an estimate of the cutoff of EFT, which is uncertain by a factor of few. This is further supported by explicit field-level calculations~\cite{Bal15,Baldauf:2015tla},
which showed that the 
matter density field 
in a field-level model equivalent to the two-loop EFT
is more than $80\%$ correlated with the
simulated non-linear field
at $k_{\rm max}^{\rm 2-loop}(z)\approx 2k_{\rm nl}(z)$. This suggests that our theory calculations at 
$k_{\rm max}^{\rm 2-loop}(z)\approx 2k_{\rm nl}(z)$
are not 
in the overfitting regime.

Next, we find that all of quadratic  contributions in $P^{\rm 2-loop}_{\rm ctr.}$ are highly degenerate, with individual error-bars on $\{e_1,e_2,e_3\}$ to be
$\mathcal{O}(200)$ at all redshifts, which violates the perturbativity 
of EFT. This implies that these parameters
(which are expected to be $\mathcal{O}(1)$ numbers, given our scalings where 
the cutoff of the EFT is on the order of 
$k_{\rm nl}(z)$) cannot be separately detected 
in practice from the power spectrum data.
Setting $e_1,e_2$ to zero, we find that the strongest remaining degeneracy is between $c_{\rm stoch}$ and $\{c^2_s,e_0,c_4,e_1\}$, which exceeds $80\%$ for all the parameters. Marginalizing over the stochasticity in the power spectrum analysis yields constraints on other EFT parameters of around $\mathcal{O}(10)$, which again clashes with the perturbative nature of EFT. Having removed $c_{\rm stoch}$
we find that, even though the remaining correlations between EFT parameters remain high, about $70\%$,
they can all be determined from our mock data with typical $\mathcal{O}(1)$
errors at each redshift. This fixes our fiducial modeling choice: we keep only the counterterms $\{c^2_s,c^2_{s,1}\equiv e_0,c_4,c_{\rm quad}\equiv e_1\}$.
With this choice we can reproduce the 
mock dark matter power spectrum to $0.5\%$ up to $k^{\rm 1-loop}_{\rm max}=2k_{\rm nl}(z)$, see Figure~\ref{fig:aem_fits}, where we also display the one-loop results at the same $0.5\%$ accuracy target. We  emphasize that reducing the
original parameter space does not
worsen the reach or the accuracy because the extra parameters are very degenerate with the
minimal set $\{c^2_s,c^2_{s,1},c_4,c_{\rm quad}\}$
that can be well constrained by data.

We note that even this set is somewhat excessive since the effective volume of our fiducial dark matter survey is quite large. If one were to rescale 
the effective volume to 
match the $S_8$ errorbars from actual weak lensing analyses (which are quite poor because the cosmic shear data is two-dimensional), 
one would realistically need only one counterterm~\cite{Baldauf:2015aha}: 
\be 
P_{\rm ctr-minimal}=-c_s^2(z)
\frac{k^2}{k_{\rm nl}^2}(P^{\rm lin}+P^{\rm 1-loop})\,,
\ee 
which reproduces the data up to $k_{\rm max}^{\rm 2-loop}$ with $\simeq 1\%$ precision at all redshifts. 
In what follows we will proceed with the non-minimal four-parameter model in order to be general.

\section{The two-loop EFT 
implementation for weak lensing analyses}

\begin{figure}[htb!]
    \centering
    \includegraphics[width=1.0\linewidth]{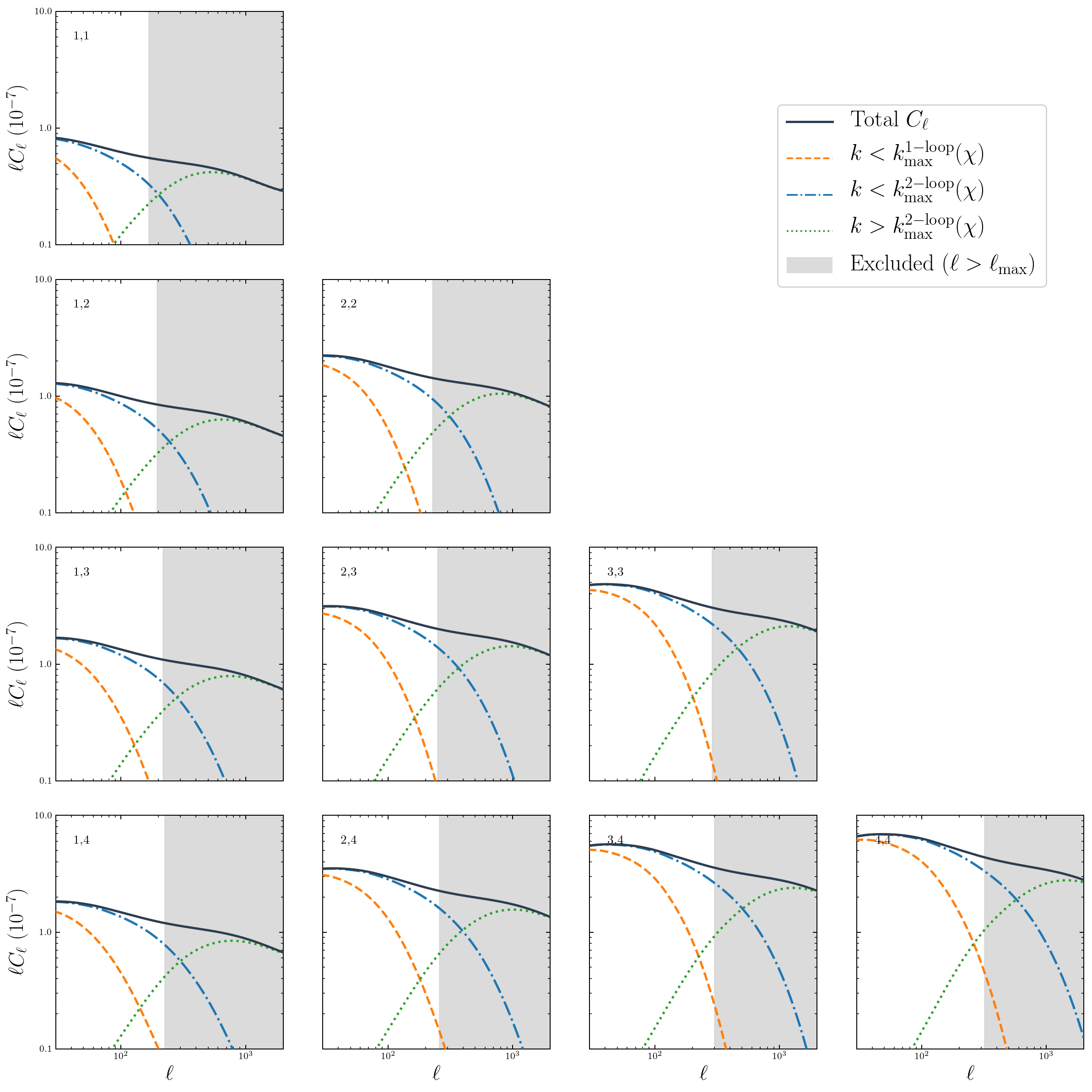}
    \caption{Decomposition of the weak lensing convergence power spectrum $C_\ell^{\kappa\kappa}$
    for all ten tomographic bin pairs of DES-Y3, as a function of multipole $\ell$. The total signal (solid dark blue) is split into contributions from three wavenumber domains evaluated at each comoving distance $\chi$ along the line of sight via the Limber approximation $k(\ell + \tfrac{1}{2})/\chi$: modes below the one-loop validity scale $k < k_{\rm max}^{\rm 1\text{-}loop}(\chi)$ (orange dashed), modes in the two-loop regime  $k_{\rm max}^{\rm 1\text{-}loop}(\chi) < k < k_{\rm max}^{\rm 2\text{-}loop}(\chi)$ (blue dot-dashed), and modes beyond the two-loop validity scale $k > k_{\rm max}^{\rm 2\text{-}loop}(\chi)$ (green dotted). The sliding scales $k_{\rm max}(\chi)$ are taken from the EFT fits to $\textsc{Aemulus}-\nu$ power spectrum emulator data in 3D. The light gray shaded region marks the conservative scale cut $\ell > \ell_{\rm max}$ applied in this analysis. Panels are labeled $(i,j)$
    for source bins $i \leq j$, with higher-redshift bins toward the bottom-right. The two-loop EFT regime dominates the signal within the analysis range for all cross-correlations involving high-redshift bins.}
    \label{fig:decompos}
\end{figure}

\textbf{Contributions to the lensing signal.}
Having determined the
momentum reach of EFT, let us estimate what fraction of the weak lensing signal can it describe. To this end, we compute the lensing convergence power
spectra in harmonic space
for the DES-Y3 source bins
using the non-linear power spectrum models from $\textsc{Aemulus}-\nu$, but split this power into contributions from one- and two-loop modes (according to the above scale-cuts). 
The results are displayed in Figure~\ref{fig:decompos}.

First, we see that the one-loop EFT predictions evaluated at $k_{\rm max}^{\rm 1-loop}(z)=0.2k_{\rm nl}(z)$ do not describe the weak lensing data well on any 
angular scale, justifying the two-loop calculation performed in this work. 
Second, we see that the extended range of two-loop EFT 
allows one to capture the lensing signal up to $\ell\sim\text{few} \times 100$. 
The gray shaded area marks the range where the non-perturbative modes with $k>k_{\rm max}^{\rm 2-loop}$ start to contribute significantly to the 
cosmological constraints. The corresponding $\ell_{\rm max}(z)$ 
is discussed below; this
closely matches that used in an official analysis of the DES-Y3 cosmic shear
in harmonic space~\cite{DES:2022qpf}. All in all, this indicates that the weak lensing data can be analyzed in EFT without significant loss of constraining power in comparison with the standard simulation-based analyses.

\begin{figure}
    \centering
    \includegraphics[width=1.0\linewidth]{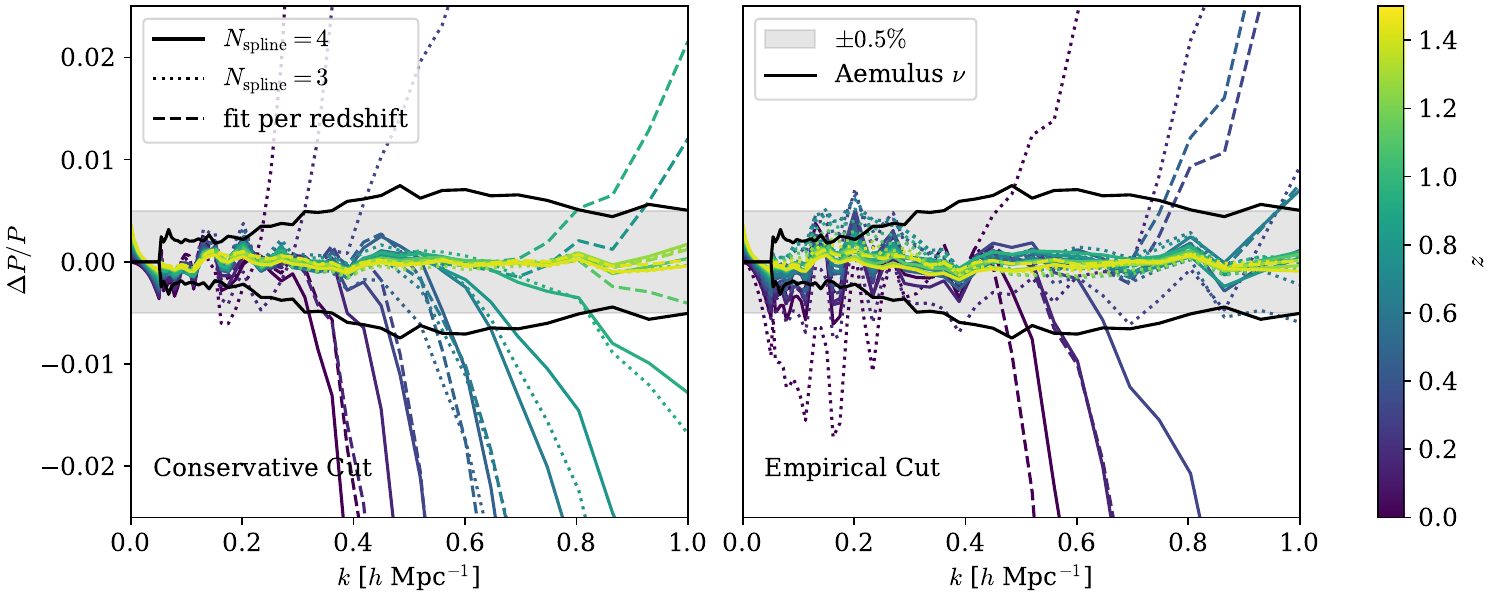}
    \caption{Fits to the matter power spectrum for Conservative ($< k_{\rm nl}(z)$) and Empirical ($< k^{\rm 2-loop}_{\rm max}(z)$) scale cuts varying the number of spline nodes in the counterterm time evolution model (Eq.~\ref{eqn:spline_time_dep}). In both scenarios, the four-node spline (solid) is able to match the performance of per-redshift fits (dashed) to the power spectrum, while the three-node spline performs substantially worse. In particular, for conservative scale cuts the four-node spline predictions nowhere differ from the non-linear matter power spectrum by more than the emulator error (shown in black). }
    \label{fig:pmm_fits}
\end{figure}

\textbf{Time-dependence of two-loop counterterms and scale cuts.}
The next important question for the weak lensing analysis is the time-dependence of the counterterms. This question is also tied to our choice of scale cuts, since in principle assuming a smoother or more restrictive time evolution can lead to worse fits than if power spectra are fit at independent redshifts.
In this section we will explore the modeling of EFT counterterms with redshift spine functions as in Eq.~\eqref{eqn:spline_time_dep}. The maximum reach of these spline function is set by the highest-redshift probed in our analysis, in this case approximately $z = 1.5$ for DES-Y3 source galaxies. We consider two choices of scale-cut:
\begin{enumerate}
    \item Conservative Cut: We fit the matter power spectrum up to $k_{\rm max} = k_{\rm nl}(z)$
    \item Empirical Cut: We fit the power spectrum at increasing $k_{\rm max}$, iterating until the best-fit power spectrum deviates from the truth by more than $0.5\%$, roughly the error of $\textsc{Aemulus}-\nu$ at $z = 0$. 
\end{enumerate}
The Empirical Cut corresponds to using $k^{\rm 2-loop}_{\rm max}(z)$ as defined in the previous section, while the conservative cut places stronger perturbativity bounds on our range of fit. The dotted lines in Figure~\ref{fig:pmm_fits} represent fits of individual redshifts as in Figure~\ref{fig:aem_fits}. While the former scale cut leads to a somewhat better fit especially at low $k$, well-within the expected emulation error of the $\textsc{Aemulus}-\nu$ emulator, the two-loop prediction from the EFT is able to fit the matter power spectrum to within sub-percent levels in both cases.

The solid and dotted lines in Figure~\ref{fig:pmm_fits} show fits to the power spectrum assuming four- and three-parameter splines for the counterterms, fit up to $k_{\rm max}(z)$ as defined for the per-redshift fits above.
While the three-parameter spline shows noticeably worse agreement with the per-redshift fit, particularly at low $z$ on large scales, the four-parameter fit matches the matter power spectrum almost identically to the per-redshift fits. Assuming again an ideal $5 \cGpc$ survey, the $95\%$ confidence intervals of the EFT parameters in these fits constrain the time-dependent predictions of the EFT parameters within the spline model to consistently 
reproduce the non-linear power spectrum of the 
emulator to $\lesssim 0.5\%$ at all redshifts, confirming our expectations from the per-redshift fits in the previous section. In all of our fits, the EFT parameters are $\lesssim 0.5$ in units of $k_{\rm nl}(z)$ at all redshifts, suggesting that our fits are under good theoretical control. We will use these N-body measurements to set our priors below, and note that counterterm corrections at these rough sizes are somewhat larger than the expected size of baryonic corrections with characteristic scale $k_{\rm bar} = 1 h$ Mpc$^{-1}$.
Since a pure measurement of the three-dimensional matter power spectrum at the implied volume would give a constraint on the total matter clustering well-beyond the statistical power of any current or upcoming weak lensing surveys, we conclude that our EFT model with time dependence described by four spline coefficients is more than adequate for both of the scale cuts explored here, with stable qualitative behavior over a range of scale cuts. From these results, we choose hard physical cutoffs in the two-loop EFT predictions in our lensing model when the power spectrum deviates by more than $3.3\%$ from the best-fit, or about $\sigma/3$ of the expected $S_8$ constraint. Since the LCTs remove the need for a harsh angular scale cut to ward off UV effects, this is a physical scale cut similar to comparing statistical and theoretical errors when analyzing spectroscopic galaxy-survey data.

\textbf{Baryonic effects and intrinsic alignments.}
EFT also provides models for the main
contaminants of cosmic shear: baryonic feedback and IA. 
For the baryons, a very accurate 
($\lesssim 0.5\%$) description is provided by the Pad\'e
resummed version of the EFT gradient 
expansion \cite{Chen24,DeRose25}
\be 
\label{eq:bars}
\frac{P_{\rm m}^{\rm bar}}{P_{\rm m}}=1-\frac{a_2(z) k^2}{1+ b_2(z)k^2}\,.
\ee 
One can check that it recovers the two-loop
EFT prediction for baryons 
in the limit $k\to  0$~\cite{Lewandowski15,Braganca:2020nhv},
\be 
\label{eq:bars2}
\frac{P_{\rm m}^{\rm bar,EFT}}{P_{\rm m}}=1-a_2 k^2 + a_4 k^4+...\,.
\ee
In this approximation, however, the baryonic corrections and the EFT DM counterterms
are completely degenerate, implying that the counterterms included above already account for the baryons. On the other hand, while the Pad\'e expression assumes that modifications to the dark-matter power spectrum can be captured purely as a transfer function, terms in the two-loop EFT such as $\delta^{\rm ctr.}_2$ also capture mode-coupling modifications due to the back-reaction of small scale modes, which should be induced by baryons to a comparable degree.

To test our baryonic modeling assumptions, we re-perform our fiducial analysis of DES-Y3 shear data replacing 
the EFT operators $k^2P^{\rm lin}$
and $k^4P^{\rm lin}$ with baryonic correction~\eqref{eq:bars}.
Using the same model for the time dependence of $a_2$ and $b_4$ (linear splines with $N=4$), we recover identical results, which confirms that the EFT dark matter model correctly accounts for baryonic effects at the level of the data.

We model IA using the 
Lagangian one-loop
EFT prediction implemented in the \texttt{spinosaurus} code~\cite{Chen23}, varying biases up to quadratic order along with the requisite EFT parameters, as in~\cite{DeRose25}. Since these are a subdominant source of systematics, we model their time-dependence via linear ($s=1$) splines with $N=2$. This choice also matches the per-parameter degrees of freedom used in standard analyses like DES-Y3~\cite{DES:2021vln}, though will likely need to be revisited for more-constraining future data, using larger $N$ to account for redshift-dependent IA contamination~\cite{DeRose26}.

\begin{figure}
    \centering
    \includegraphics[width=0.9\linewidth]{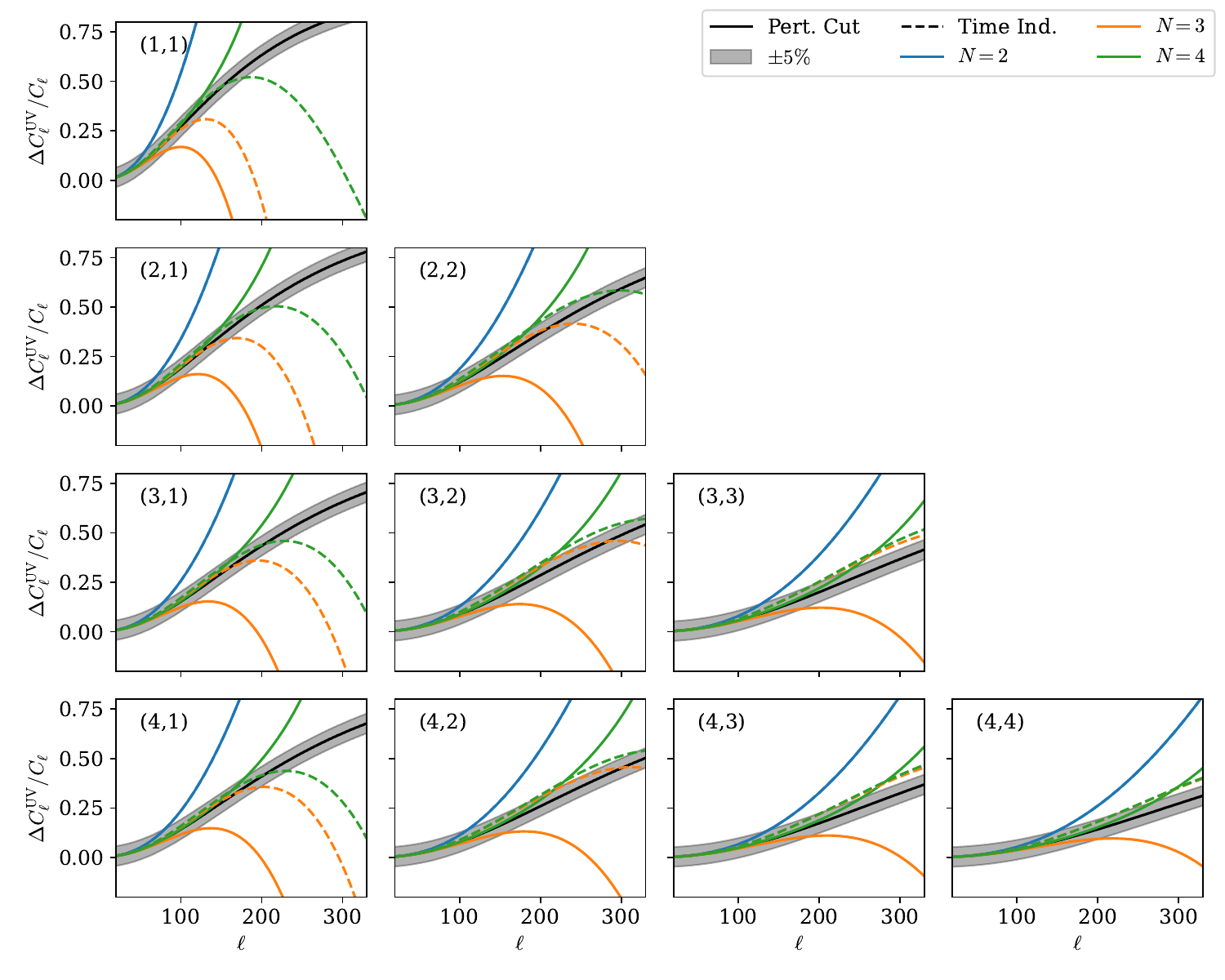}
    \caption{Convergence of the lensing counterterm expansion using the Conservative cuts, compared to the full UV contribution to the lensing integral in a scenario where the true non-linear matter power spectrum is given by $\textsc{Aemulus}-\nu$. The LCT expansion recovers the lensing signal to within e.g. $5\%$ (gray bands) even when the total UV contribution is a significantly larger fraction of the signal. Dashed lines show the same expansion when the LCTs are calculated without accounting for the time dependence of the cutoff $\Lambda$, and show less asymptotic agreement with the black curve in all cases.}
    \label{fig:lct_expansion}
\end{figure}

\begin{figure}[htb!]
    \centering
    \includegraphics[width=1.0\linewidth]{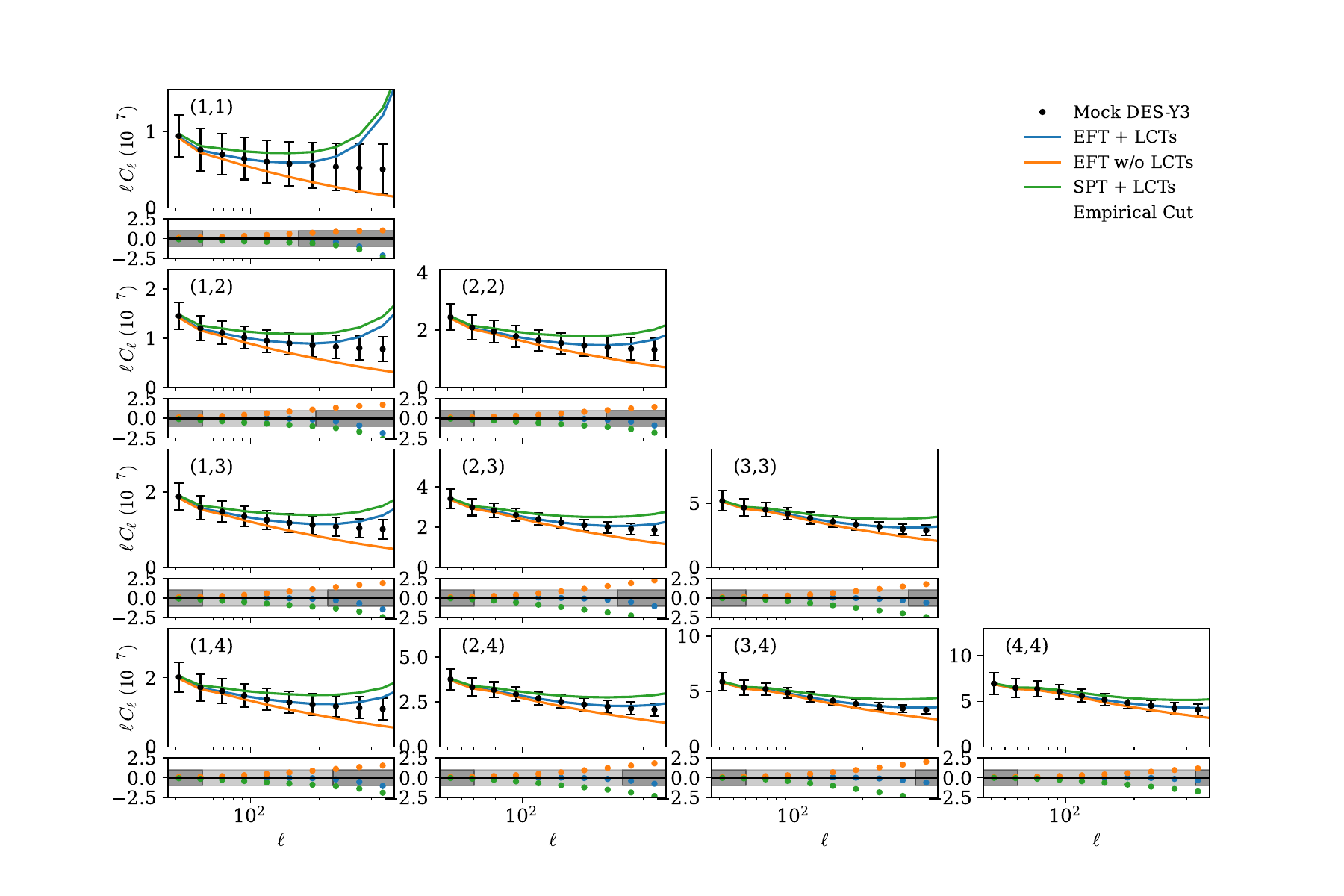}
    \caption{Comparison of two-loop predictions for the DES-Y3 cosmic shear data (blue) compared to non-linear mocks (black points). The orange and green curves show the model prediction without lensing counterterms or EFT counterterms, such that the lensing integrals and loop integrals are respectively un-renormalized. We see that in either case these sub-models are unable to recover the expected lensing signal within the true cosmology, unlike the full prediction.}
    \label{fig:model_validation}
\end{figure}
\begin{figure}
    \centering
    \includegraphics[width=0.5\linewidth]{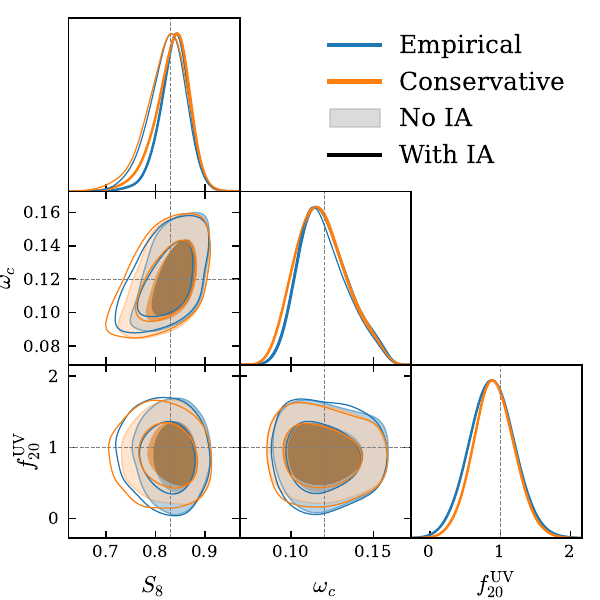}
    \caption{Fits to the cosmological parameters using the mock data in Figure~\ref{fig:model_validation} in the empirical and conservative cuts (blue and orange), with and without marginalizing over galaxy intrinsic alignments (unfilled and solid). In all cases the posteriors are able to well-recover the true cosmology.}
    \label{fig:model_validation_fit}
\end{figure}
\begin{figure}[htb!]
    \centering
    \includegraphics[width=1.0\linewidth]{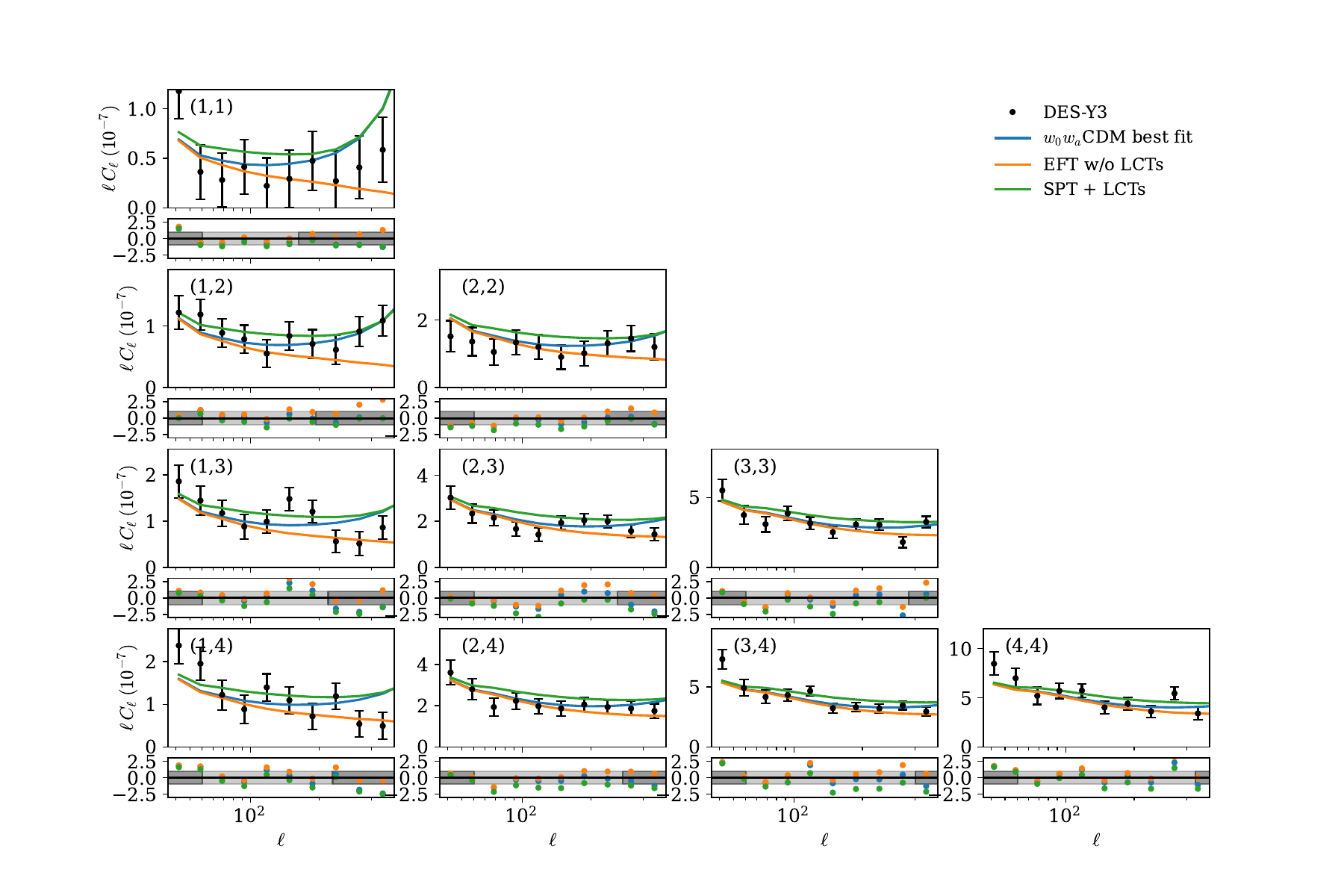}
    \caption{Similar to Figure~\ref{fig:model_validation} but applied to the DES-Y3 data.}
    \label{fig:best_fit_data}
\end{figure}

\begin{figure}
    \centering
    \includegraphics[width=0.99\linewidth]{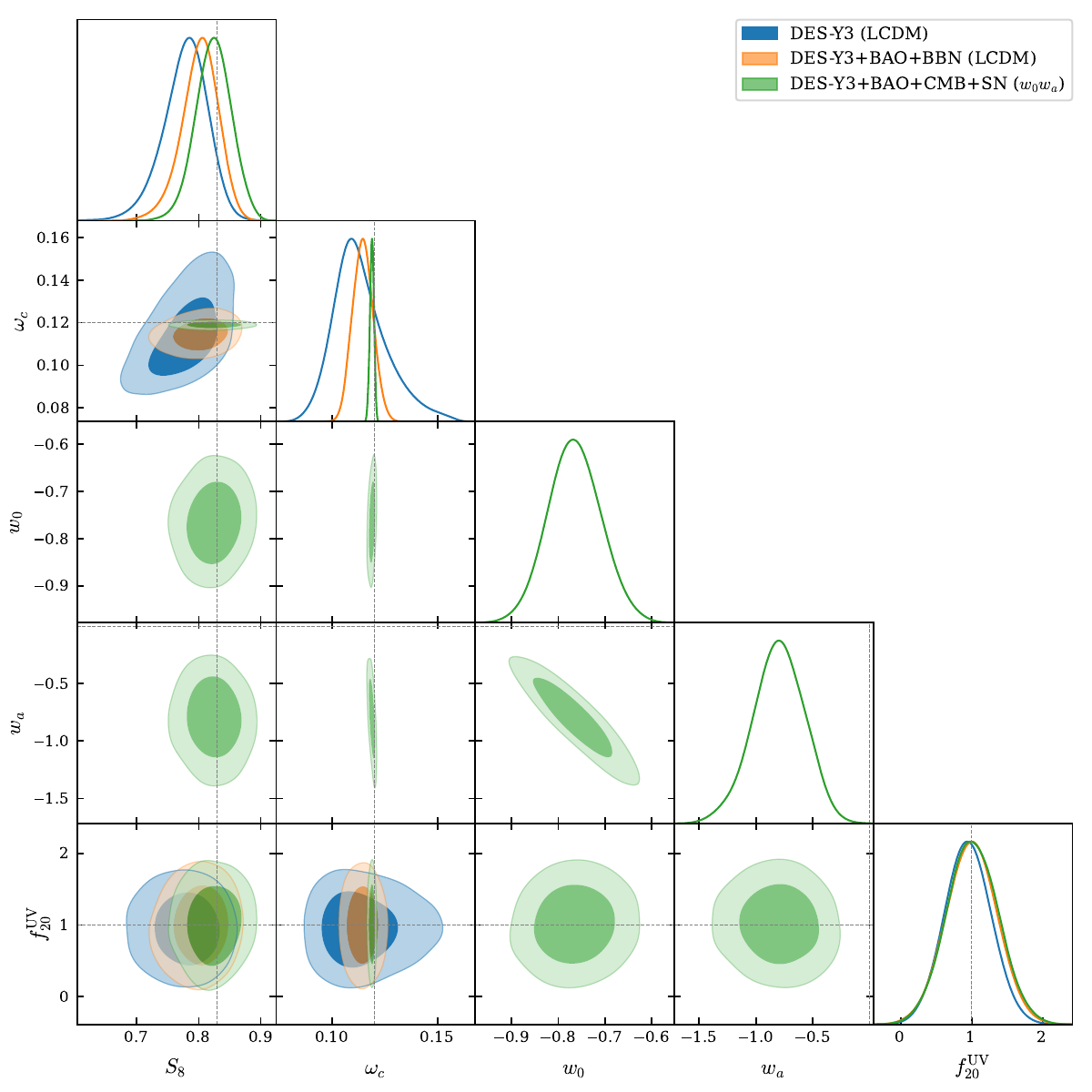}
    \caption{Cosmological constraints from the DES-Y3 data including the dark energy parameters.}
    \label{fig:full_constraints}
\end{figure}

\textbf{Lensing counterterms.} Ref.~\cite{DeRose25} showed that the UV sensitivities of the lensing 2-point function can be absorbed by marginalizing over lensing counterterms (LCTs)
\begin{equation}
    C_\ell^{\kappa_i \kappa_j} \supset \sum_{N=2}^\infty (\ell + 1/2)^{N-1} \sum_{n+m+o+N} w^i_n w^j_m \sigma_{N,o}
\end{equation}
where $w^i_n$ are the n$^{\rm th}$ order Taylor-series coefficients of the lensing kernel of galaxy sample $i$ in terms of the comoving distance $\chi$. The LCTs describe the small-scale matter power spectrum and are given by the integrals
\begin{equation}
        \sigma_{N,o} = \frac{1}{o!} \int dk \ k^{-N}\ \left( \frac{d^o P_{\rm UV}}{d\chi^o} \right)_{z=0}
\end{equation}
where $P_{\rm UV}(k,\chi)$ is the small-scale part of the non-linear matter power spectrum at wavenumber $k$ and comoving distance $\chi$. For example, in the case of a hard cutoff at $\Lambda$, the lensing counterterms simply involve an integral $\int_{k > \Lambda} dk$. However, the form of the LCT expansion is general and not specific to any given method of regularizing the lensing integral (Eqn.~\ref{eq:Cell}).

Of particular interest to perturbative predictions of LSS  is the case where the the cutoff $\Lambda = \Lambda(\chi)$ is time dependent, since the fluid description underlying the EFT breaks down on larger scales at later times. Here, we show that the lensing counterterm expansion applies also in this scenario and estimate the size of the contributions. We first define the time-dependent UV contribution piece
\begin{equation}
    P_{\rm UV}^{(t)}(k,\chi) = P(k,\chi) \Theta(k - \Lambda(\chi)) - P(k,\chi) \Theta(k - \Lambda_0) 
\end{equation}
where $\Lambda_0 = \Lambda(0)$ is the theory cutoff at $z=0$. In this case, we can write the change in lensing counterterms due to time dependence as
\begin{equation}
    \sigma_{N,o} = \sigma_{N,o}^{(t)} + \sigma_{N,o}^{(0)}
\end{equation}
where $\sigma_{N,o}^{(0)}$ is the LCT evaluated for a fixed cutoff $\Lambda^{(0)}$ and $\sigma^{(t)}_{N,o}$ is given by
\begin{equation}
    \sigma^{(t)}_{N,o} = \frac{1}{o!} \int dk \ k^{-N}\ \left( \frac{d^o P_{\rm UV}^{(t)}}{d\chi^o} \right)_{z=0}.
    \label{eqn:lcts}
\end{equation}
The time-dependent cutoff in the Heaviside function leads to Dirac deltas upon differentiation, leading the counterterms to become functions of the power spectrum and its derivatives evaluated at $\Lambda^{(0)}$. The first few nonzero contributions are
\begin{align}
    \sigma^{(t)}_{3,1} &= -\frac{\Lambda '(0) P(\Lambda^{(0)},0)}{\left(\Lambda^{(0)}\right)^3}, \quad  \sigma^{(t)}_{4,1} = -\frac{\Lambda '(0) P(\Lambda^{(0)},0)}{\left(\Lambda^{(0)}\right)^4} \nonumber \\
   \sigma^{(t)}_{4,2} &= \frac{P(\Lambda^{(0)},0) \left(4 \Lambda'(0)^2-\Lambda^{(0)} \Lambda''(0)\right)-\Lambda^{(0)} \Lambda'(0) \left(\Lambda'(0) \
P^{(1,0)}(\Lambda^{(0)},0)+2 P^{(0,1)}(\Lambda^{(0)},0)\right)}{2 (\Lambda^{(0)})^5}
\label{eqn:lct_time_dep}
\end{align}
The first two terms describe the entry of modes at $\Lambda^{(0)}$ into the perturbative regime at $z > 0$, while the time and wavenumber derivatives of the power spectrum in the second line describe modulations due to the relative slope of the power spectrum and its time evolution. In this work we will use the LCT expansion up to third order in $\ell$; the expansion has a radius of convergence defined by the mean separation to background source galaxies, and Ref.~\cite{DeRose25} showed that going to higher $\ell^n$ did not yield significant improvements.

Figure~\ref{fig:lct_expansion} shows the convergence of the LCT expansion for DES Y3 sources in the case of the perturbative cut. In order to validate the convergence of expansion, we have computed the ``true'' UV contribution of modes beyond the redshift-dependent cutoff using the fully non-linear matter power spectrum predicted using the $\textsc{Aemulus}-\nu$ emulator, as shown via the solid black lines bounded by a $5\%$ error window. In this case, since both the UV and IR theories are completely known, we can compute the lensing counterterms exactly following Equation~\ref{eqn:lcts} and \ref{eqn:lct_time_dep}. The time-dependent LCT expansion, shown in the solid lines, evidently converges upon the truth, with agreement better at low $\ell$ and higher $N$. The LCTs further better match the truth when the source galaxies are at higher redshift, which is as expected given that it relies on the smoothness of the power spectrum and lensing kernels at low $z$. For comparison, the dashed lines also show the LCT expansion computed without time-dependent terms. This expansion is equivalent at leading order, where only the total missing power above the cutoff $\Lambda^{(0)}$ matters, but the solid lines show better agreement at higher $\ell$ that is particularly apparent at $N=4$ when the range of applicability of the expansion is wider.

The LCT expansion offers a pragmatic way to define angular scale cuts, $\ell_{\rm max}(z)$, given physical scale cuts, $k_{\rm max}(z)$. For a fiducial dark matter power spectrum, we can compute the ``true'' LCTs and UV contributions in the fiducial cosmology and evaluate where the expansion breaks down. In practice, since we expect $\sim 5\%$ precision on $S_8$, or $\sim 10\%$ constraints on the lensing power spectrum amplitude, we can drop points that deviate by more than one third of a sigma at $N=4$ in the LCT expansion. For the Conservative and Empirical Cuts, this gives
\begin{align}
    \ell_{\rm max}^{\rm Conservative} &= [125, 148, 174, 165, 198, 226, 174, 205, 233, 249] \nonumber \\
    \ell_{\rm max}^{\rm Empirical} &= [167, 195, 228, 216, 249, 287, 224, 258, 301, 319]
\end{align}
where we set $\ell_{\rm min} = 50$ in all cases to avoid beyond-Limber contributions. We note that these cuts are in themselves rather conservative, since the theoretical error in the LCT expansion they correspond to become comparable to the statistical errors in our analysis only very close to the smallest angular scale fit.

\section{Analysis of DES-Y3 cosmic shear}

\textbf{Details of the Analysis.} We adopt priors $ A_{\sigma_8}^n c^{(n) i}_O \in \mathcal{N}(0, 0.5^2)$ on all the EFT counterterms in our analysis, where the factor of $A_{\sigma_8} = \sigma_8(z) / \sigma_{8, \rm fid}(z)$ roughly fixes the contribution of each counterterm as a function of cosmology.  Since the spline basis forms a partition of unity this also sets the overall magnitude of $c_O(z).$ We similarly sample over the EFT coefficients for intrinsic alignments for spin-2 operators up to quadratic order 
\begin{equation}
    g_{ij}(\textbf{q}) = A_1 s_{ij} + A_{\delta 1} \delta s_{ij} + A_2 \text{TF}\{ s^2_{ij} \} + A_t t_{ij} + \alpha_s \nabla^2 s_{ij} + \epsilon_{ij}
\end{equation}
where the coefficients $A_O$ are normalized as is conventional in the literature by $ \sigma_8^n (z)A_{O^{(n)}} = c_{O^{(n)}} \bar{C}_1 \rho_c \Omega_{m,\rm fid} \sigma_{8, \rm fid}^n$ and we sample with prior $c_{O^{(n)}} \in \mathcal{N}(0,5^2)$ based on conservative assumptions about the intrinsic alignments of halos measured in simulations \cite{Akitsu23}. We set the prior on derivative operator coefficient $\alpha_s$ by $\mathcal{N}(0, (0.5 \times 5 k_{\rm IA}^{-2})^2)$, where $k_{\rm IA} =0.4\ h$ Mpc$^{-1}$, i.e such that its contribution can be at most about $50\%$ of the linear theory signal at that scale. We fix the stochastic contribution $\epsilon_{ij}$ to have variance equal to the shape noise. These assumptions are described in more detail in refs.~\cite{Chen:2024vvk,DeRose25}. We define the parameters
\begin{equation}
    f_{N,o}^{\rm UV} = \frac{\sigma_{N,o}}{\sigma_{N,o}^{\rm dmo}}
\end{equation}
where $\sigma_{N,o}^{\rm dmo}$ is the lensing counterterm assuming the true UV theory is given by dark-matter only N-body simulations, in this case the $\textsc{Aemulus}$ $\nu$ at each cosmology. We assume $f^{\rm UV}_{N,o} \in \mathcal{N}(1, 0.4^2)$ to cover a wide range of hydrodynamical simulations \cite{DeRose25}.

Throughout the analysis we use the cosmological parameter priors
\begin{equation}
    \omega_c \in \mathcal{U}(0.08, 0.16), \quad h \in \mathcal{U}(0.52, 0.82), \quad A_s \in \mathcal{U}(1.1\times 10^{-9},3.1\times 10^{-9})
\end{equation}
while setting $\omega_b = 0.02237$ and $n_s = 0.9649$. These priors are set by the parameter boundaries of the $\textsc{Aemulus}-\nu$ suite, which we use to estimate the size of LCTs, and are sufficient given our LCDM analysis mostly constrains $S_8$ while our extended model fits constrain $\omega_c$ and $h$ to a much narrower range, though they can be easily extended if fixing the cosmology dependence of the LCTs to be cosmology independent.

In addition to the cosmological and EFT/IA  parameters, we also marginalize over the 
shear calibration and redshift uncertainties 
following the standard approach adopted for DES-Y3,
see~\cite{DeRose25} for more detail. Our statistical inference was run via the Metropolis-Hastings algorithm implemented in \texttt{gholax}\footnote{https://github.com/j-dr/gholax/} using \textsc{BlackJax} \cite{blackjax}.

\textbf{Tests on non-linear mocks.} As a first test, we fit the matter spectrum directly in simulations in three dimensions as in Fig.~\ref{fig:pmm_fits} and input the resulting EFT counterterms into our theoretical model for weak lensing, along with LCTs computed as in the previous section. Figure~\ref{fig:model_validation} shows the results---evidently, our EFT model can describe the weak lensing data with excellent precision out to our chosen scale cuts. For comparison we also show two variants of the model predictions (1) the EFT prediction without LCTs to renormalize the short-scale dependence of the lensing integral, shown in orange and (2) the prediction using SPT (\textit{i.e.}\ perturbation theory with no counterterms) but including LCTs. It is apparent that both of these sub-models fail to capture the weak lensing power spectrum on typical scales used in weak lensing analyses. The LCTs are especially significant at low redshifts, indicating the weight of the short modes in the lensing integral in that regime.

We next test whether our pipeline is able to recover the true cosmology when fitting this mock data. Figure~\ref{fig:model_validation_fit} shows the fit when using both the Conservative and Empirical cuts when including or omitting intrinsic alignments. Unlike for real data, these mock data correspond to the lensing signal in pure dark matter simulations and so are uncontaminated by IA; the latter fits therefore directly test whether our dynamical model for structure formation is sufficiently accurate. The la In both cases we are able to recover the true value of $S_8$ to well within the statistical requirements of the data, with the mean within the truth at $< \sigma / 3$ in all cases. The slight differences are indeed most likely due to projection and prior volume effects, given that the ``truth'' as shown above in Figure~\ref{fig:model_validation} gives a near-perfect fit to the data---this conclusion is also borne out by examining the best-fit elements of each chain. The situation is similar for the first lensing counterterm $f^{\rm UV}_{20}$. We note that the dark-matter density $\omega_c$ is also well-recovered by the analysis, though the resulting constraints are not particularly strong and would therefore be superceded by external priors from e.g. the CMB in realistic analyses.

\textbf{Additional information on fits to DES-Y3 data.} Figures~\ref{fig:best_fit_data} shows the best fit $w_0 w_a$CDM prediction using our EFT model, highlighting the importance of the EFT counterterms and LCTs as in Figure~\ref{fig:model_validation}. Expanding on Figure~\ref{fig:constraints}, the cosmological posteriors of our fits including the dark-energy parameters $w_0 w_a$ are shown in Figure~\ref{fig:full_constraints}.

\end{document}